\documentclass[a4paper]{article}
\usepackage{command_MU}
\usepackage{algorithm, algpseudocode} 
\usepackage{mathrsfs} 

\usepackage{amsmath}
\usepackage{amsthm}
\usepackage{amssymb}
\usepackage[shortlabels]{enumitem}
\usepackage{stfloats}
\usepackage{siunitx}
\usepackage[utf8]{inputenc}
\usepackage{booktabs}
\usepackage{hyperref}
\hypersetup{
	unicode,
	pdfauthor={Author One, Author Two, Author Three},
	pdftitle={A simple article template},
	pdfsubject={A simple article template},
	pdfkeywords={article, template, simple},
	pdfproducer={LaTeX},
	pdfcreator={pdflatex}
}
\usepackage[sort&compress,numbers,square]{natbib}
\theoremstyle{plain}

\theoremstyle{definition}
\newtheorem{remark}{Remark}

\usepackage{graphicx, color}
\graphicspath{{fig/}}

\newcommand{\fixme}[1]{\textcolor{black}{#1}}
\DeclareMathOperator{\prox}{prox}

\title{Iteratively Refined Image Reconstruction\\ with Learned Attentive Regularizers}
\author{Mehrsa Pourya$^{\dagger}$, Sebastian Neumayer$^{\ddagger}$, and Michael Unser$^{\dagger}$ \thanks{mehrsa.pourya@epfl.ch, sebastian.neumayer@math.tu-chemnitz.de, michael.unser@epfl.ch}}

\date{
	$\dagger$ Biomedical Imaging Group, EPFL\\ \vspace{1.5mm} 
        $\ddagger$ Professorship of Inverse Problems, TU Chemnitz
}

\begin{document}
	\maketitle
	\begin{abstract}
          We propose a regularization scheme for image reconstruction that leverages the power of deep learning while hinging on classic sparsity-promoting models.  
          Many deep-learning-based models are hard to interpret and cumbersome to analyze theoretically.
           In contrast, our scheme is interpretable because it corresponds to the minimization of a series of convex problems. 
          For each problem in the series, a mask is generated based on the previous solution to refine the regularization strength spatially.
          In this way, the model becomes progressively attentive to the image structure.
          For the underlying update operator, we prove the existence of a fixed point.
          As a special case, we investigate a mask generator for which the fixed-point iterations converge to \fixme{a critical point} of an explicit energy functional.
          In our experiments, we match the performance of state-of-the-art learned variational models for the solution of inverse problems.
          Additionally, we offer a promising balance between interpretability, theoretical guarantees, reliability, and performance.

	\end{abstract}
	
\section{Introduction}\label{sec:Intro}
  In biomedical imaging \cite{MM2019}, including magnetic resonance imaging (MRI) and computed tomography, reconstructions are often achieved via the resolution of an inverse problem.
  Its task is to recover an unknown signal $\V x \in \R^N$ from noisy measurements $\V y = \M H \V x + \V n \in \R^M$, where $\M H \in \R^{M, N}$ encodes the data-acquisition process and the noise $\V n \in \R^M$ accounts for imperfections in this description.
  From a variational perspective \cite{Scherzer2009}, one defines the reconstruction as the solution to the minimization problem 
    \begin{equation}
        \label{eq:dis_inv_pro}
        \argmin_{\V x \in \R^N} \bigl(\Op{E}(\M H \V x, \V y) + \lambda \mathcal{R} (\V x) \bigr),
    \end{equation}
    which involves a data-fidelity term $\Op{E} \colon \R^M \times \R^M \to \R_{\geq 0}$ and a regularizer $\mathcal{R} \colon \R^N \to \R_{\geq 0}$.
    In \eqref{eq:dis_inv_pro}, the data fidelity term ensures the consistency of the reconstruction with the measurements, while the regularization, whose strength is controlled by $\lambda \in \R_{>0}$, imposes some regularity constraints (prior information) on the solution.

   For a large variety of data-acquisition and noise models, a well-studied zoo of data fidelities $\Op{E}$ can be found in the literature.
   While an instance-specific $\Op{E}$ is natural, it is desirable that the regularizer $\mathcal{R}$ is agnostic to $\M H$ and $\V n$ and solely depends on the properties of the underlying images.
   Hence, a regularizer that captures these inherent properties would be of great interest.
   Attempts can be traced as far back as to the Tikhonov regularization \cite{tikhonov1963}, where images are modeled as smooth signals.
   Later, this approach was outperformed by compressed sensing \cite{donoho2006compressed}.
   Such models either assume that the signal is sparse in some latent space (e.g., wavelet decomposition \cite{Ma89}) or involve a filter-based regularizer $\mathcal{R}$ such as the total variation (TV) \cite{rudin1992nonlinear, donoho2006compressed} and its generalizations \cite{lefkimmiatis2011hessian}.
   These classic signal-processing approaches achieve a baseline performance with the advantage that they provide stability and robustness guarantees \cite{PlaNeuUns2023}. 
   
   With the emergence of deep-learning techniques for the solution of inverse problems \cite{AMOS2019}, the traditional approaches have been outperformed in many applications.
   The end-to-end training achieves state-of-the-art performance in terms of quantitative metrics such as the peak signal-to-noise ratio (PSNR).
   However, such models are often neither interpretable nor trustworthy for sensitive applications such as biomedical imaging \cite{zbontar2018fastMRI,antun2020instabilities}. 
   Therefore, a recent line of research \cite{LiSch2020,lunz2018adversarial,KobEff2020,DufCamEhr2023} is focusing on the use of deep learning for the solution of inverse problems within the variational framework \eqref{eq:dis_inv_pro}.
   There, instead of learning the whole reconstruction pipeline in an end-to-end manner, one only learns the regularizer $\mathcal{R}$.
   Up to now, these models have relied mostly on deep architectures to parameterize $\mathcal R$, which makes an interpretation difficult.
   To bypass this issue, the authors in \cite{GouNeuBoh2022} have proposed to  parameterize the learnable $\mathcal{R}$ as 
    \begin{equation}
        \label{eq:cprp_reg}
        \mathcal{R}\colon\V{x}\mapsto \sum_{c=1}^{ N_\mathrm{C}} \bigl \langle \mathbf{1}_{N}, \boldsymbol \psi_c( \M W_{c} \V x) \bigr \rangle,
    \end{equation} 
    with channel-wise data-driven convolutional matrices $\M W_{c} \in \R^{N, N}$ and $\boldsymbol \psi_c (\V x) = (\psi_c(x_k))_{k=1}^N$, where the convex and symmetric profiles $\psi_c$ are members of \smash{$\mathcal{C}_{\geq0}^{1,1}(\R)$}, the space of nonnegative differentiable functions with Lipschitz-continuous derivatives.
    Based on the architecture \eqref{eq:cprp_reg}, the authors of \cite{GouNeuBoh2022} obtain the best performance among known convex regularizers in their experiments.
    Moreover, \eqref{eq:cprp_reg} has a clear interpretation as a filter-based regularizer.
    To further improve the reconstruction performance, we need to look beyond convexity.
    As an extension of the model \eqref{eq:cprp_reg}, the authors of \cite{GouNeuUns2023} have proposed to learn symmetric potentials \smash{$\psi_c \in \mathcal{C}_{\geq 0}^{1,1}(\R)$} with $\psi_{c}^{\prime \prime} \geq -\rho$ a.e., namely $\rho$-weakly convex ones.
    This relaxation significantly improves over the convex setting.
    In particular, it gets close to the performance of the DRUNet-based model \cite{HurLec2022}, which is among the best-performing methods with a (loose) energy interpretation.
    \begin{figure}[t]
    \begin{center}
      \includegraphics[width=15cm,height=10cm,keepaspectratio]{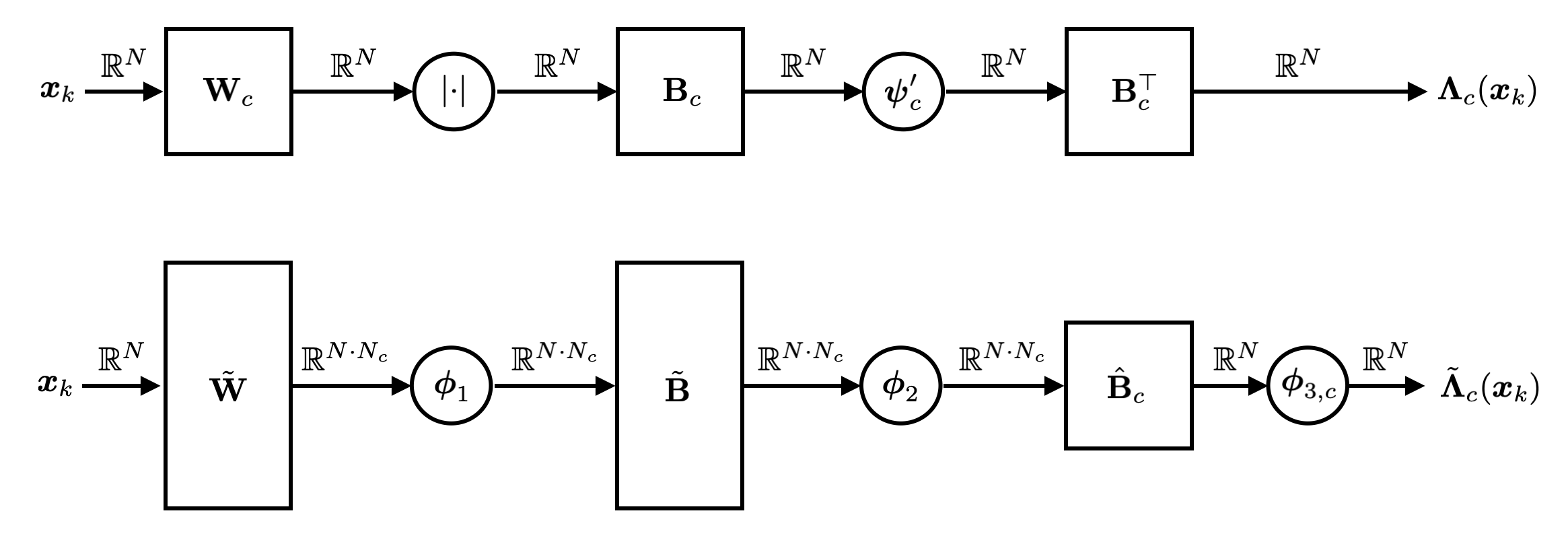}
    
      \caption{Mask-generation architecture of the majorization-minimization (top) and the solution-driven (below) setting.
      Above each arrow, we denote the signal dimension at the corresponding stage.}
      \label{fig:freu_box}
    \end{center}
    \end{figure}
    \paragraph{Outline and Contribution}
    First, we introduce the theoretical concepts in Section~\ref{sec:preliminaries}.
    Then, we establish in Section~\ref{sec:MajMin} a link between the use of a $\rho$-weakly convex $\psi_{c}$ within \eqref{eq:cprp_reg} and spatially-adaptive regularization \cite{HiPaRa2017,ChReSc2017,lefkimmiatis2023learning,KofAltBa2023}.
    To this end, we investigate the regularizer 
    \begin{equation}
        \label{eq:prp_reg}
        \fixme{\mathcal{R}_{\text{MMR}}}\colon\V{x}\mapsto \sum_{c=1}^{ N_\mathrm{C}} \bigl \langle \mathbf{1}_{\fixme{N}}, \boldsymbol \psi_c  \big( \M B_{c}  \abs{\M W_{c} \V x}\big)\bigr\rangle, 
    \end{equation} 
    where the convolutional matrices $\M W_{c} \in \R^{N,N}$ and \smash{$\M B_{c} \in \R_{\geq 0}^{N,N}$}, and the $\boldsymbol \psi_c (\V x) = (\psi_c( x_k))_{k=1}^N$ with concave potentials \smash{$\psi_c \in \mathcal{C}_{\geq 0}^{1,1}(\R_{\geq 0})$} are data-driven.
    In \eqref{eq:prp_reg}, $\vert \cdot \vert$ is applied component-wise to the vector $\M W_{c} \V x$ and $\M B_c$ is constrained to have normalized rows.
    As shorthand, we introduce the notations $\M W  = (\M W_{c})_{c=1}^{N_C}$, $\M B  = (\M B_{c})_{c=1}^{N_C}$, and $\boldsymbol \psi  = (\boldsymbol \psi_{c})_{c=1}^{N_C}$.
    For the regularizer \eqref{eq:prp_reg}, we show in Theorem~\ref{thm:existence} that the variational problem \eqref{eq:dis_inv_pro} is guaranteed to have at least one minimizer.
    To reach the latter, we propose to use the iterative majorization-minimization regularization (MMR) characterized by
    \begin{equation}
        \label{eq:set_cvx_prblms}
        \V x_{k+1} \in \argmin_{\V x \in \R^N} \bigl( \Op{E}(\M H \V x, \V y) + \lambda \fixme{\mathcal{R}_{\text{MMR}, k}} (\V x) \bigr),
    \end{equation}
    with initialization $\V x_1 \in \R^N$ and
    \begin{equation}
        \label{eq:reg_rw}
        \fixme{\mathcal{R}_{\text{MMR}, k}} \colon\V{x}\mapsto \sum_{c=1}^{ N_\mathrm{C}} \bigl \langle \boldsymbol \Lambda_c(\V x_k), \abs{\M W_{c} \V x}\bigr\rangle,
    \end{equation} 
    where $\boldsymbol \Lambda = (\M \Lambda_c)_{c=1}^{N_C} \colon \R^N \to (\R_{\geq 0}^N)^{N_C}$ with $\boldsymbol \Lambda_c(\V x) = \M B_c^{\top} \boldsymbol \psi^{\prime}_c(\M B_c \abs{\M W_{c} \V x})$.
    If $E$ is strictly convex and differentiable, then \eqref{eq:set_cvx_prblms}, namely the majorized problem at the $k$-th step, is strictly convex.
    Its unique minimum can be computed using the forward-backward splitting (FBS) algorithm \cite{beck2009fast}. 
    Hence, we can rewrite \eqref{eq:set_cvx_prblms} using the associated update operator $T_{\M  \Lambda, \M W, \V y}\colon \R^N \to \R^N$ as
    \begin{equation}\label{eq:fixediters}
        \V x_{k+1} = T_{\M \Lambda, \M W, \V y} (\V x_k).
    \end{equation}
    In Theorem~\ref{thm:convergnece}, we prove that the iterations \eqref{eq:fixediters} converge to a critical point of the underlying problem \eqref{eq:dis_inv_pro}.
    
    In \eqref{eq:reg_rw}, we can interpret $\boldsymbol \Lambda_c$ as a channel-wise spatial adaption of the regularization strength, which is attentive to image structures.
    This viewpoint of solution-driven spatial adaptivity \cite{LenLelBec2014,LenBer2015} serves as a starting point for the generalization of the MMR model in Section~\ref{sec:SolDrivAdapt}.
    More precisely, we propose to replace $\boldsymbol \Lambda$ in \eqref{eq:reg_rw} with a more expressive convolutional neural network $\tilde{\M \Lambda}$.
    For this, we relax the constraints on the activation functions and linear operators in the mask generator $\M \Lambda$ associated with \eqref{eq:prp_reg}, see Figure~\ref{fig:freu_box}.
    This leads to \fixme{the solution-adaptive fixed-point iterations (SAFI) as reconstruction scheme, which involves the regularizers}    \begin{equation}\label{eq:RegConvMask}
        \fixme{\mathcal{R}_{\text{SAFI}, k}} \colon\V{x}\mapsto \sum_{c=1}^{ N_\mathrm{C}} \bigl \langle \tilde{\M \Lambda}_c(\V x_k), \abs{\M W_{c} \V x}\bigr\rangle.
    \end{equation} 
    \fixme{In \eqref{eq:RegConvMask}}, $\tilde{\M \Lambda} \colon \R^N \to ([0,1]^N)^{N_C}$ is a 3-layer network  with $\tilde{\M \Lambda}_c(\V x) = \boldsymbol \phi_{3, c}(\hat{\M B}_{c} {\boldsymbol \phi}_{2} (\tilde{\M B} {\boldsymbol \phi}_1 (\tilde{\M W} \V x)))$.
    Its convolutional operators have dimensions $\tilde{\M W} \in \R^{(N_C \cdot N), N}$, $\tilde{\M B} \in \R^{(N_C \cdot N), (N_C \cdot N)}$, and \smash{$\hat{\M B}_{c} \in \R^{N, (N_C \cdot N)}$}.
    The activation functions \smash{$\boldsymbol  \phi_1(\V x) = (\phi_{1, \lceil k/N \rceil }(x_k))_{k=1}^{N_C \cdot N}$} and \smash{$\boldsymbol \phi_2(\V x) = (\phi_{2, \lceil k/N \rceil }(x_k))_{k=1}^{N_C \cdot N}$} share linear splines $\phi_{r, c} \in \mathcal{C}(\R)$ on input blocks of size $N$.
    The final activation functions are $\boldsymbol \phi_{3, c}(\V x) = (\phi_{3, c}(x_k))_{k=1}^N$, where each $\phi_{3, c}\in \mathcal{C}(\R)$ is composed of a linear spline and a Sigmoid function.
    The latter enforces that the entries of each \smash{$\tilde{\M \Lambda}_c$} remain in $[0,1]$.
    For the regularizers \eqref{eq:RegConvMask}, the minimization problem \eqref{eq:set_cvx_prblms} is still strictly convex.
    Therefore, each update \eqref{eq:fixediters} in the pipeline is numerically tractable and gives rise to an update operator \smash{$T_{\tilde{\M  \Lambda}, \M W, \V y}\colon \R^N \to \R^N$}.
    In Theorem~\ref{thm:ExistFix}, we prove that $T_{\tilde{\M  \Lambda}, \M W, \V y}\colon \R^N \to \R^N$ admits at least one fixed point. 
    In this relaxed setting, the convergence of the SAFI scheme to a fixed point is encouraged by the use of regularization techniques during training \cite{AniAshKai2022}.
    The parameterization details for the architectures \eqref{eq:reg_rw} and \eqref{eq:RegConvMask} are given in Section~\ref{sec:Parameterization}.
    The learning of the associated parameters on denoising problems is discussed in Section~\ref{sec:Training}.
    
    Our numerical evaluation for both denoising and MRI reconstruction in Section~\ref{sec:NumericalResults} indicates that the learning of the parameters $\M W$, $\M B$, and $\boldsymbol \psi$ in \eqref{eq:prp_reg} leads to a reconstruction performance similar to that of the weakly convex model from \cite{GouNeuUns2023} with $\rho=1$.
    By setting $\M B_c = \textbf{Id}$ in \eqref{eq:reg_rw}, we obtain a weakly convex regularizer without a bound on $\rho$ as special case.
    Hence, our theoretical analysis, corroborated by the numerical results, leads to yet another reasonable explanation for the performance gain of the weakly convex model \cite{GouNeuUns2023} over the convex one \cite{GouNeuBoh2022}. 
    With the more general regularizer \eqref{eq:RegConvMask} associated to SAFI, the performance gets similar to that of \cite{NeuPouGou2023}, despite the much simpler mask generator \smash{$\tilde{\M \Lambda}$}.
    As in previous works, we observe that the learned regularizers generalize well to the previously unseen inverse problems.
    Finally, conclusions are drawn in Section~\ref{sec:Conclusions}.

    \paragraph{Relation to Previous Work}
    \fixme{Our regularizer $\mathcal R_{\text{MMR}}$ relies on the architecture \eqref{eq:cprp_reg} from \cite{GouNeuUns2023}, where we add the inner activation $\vert \cdot \vert$ and the nonnegative convolutional matrices $\M B_c$.
    The decomposition $\psi_c = \mu \psi_{c, \mathrm{cvx}} + \psi_{c, \mathrm{ccv}}$ is proposed in \cite{GouNeuUns2023} with a convex $\psi_{c, \mathrm{cvx}} \in \smash{\mathcal{C}_{\geq 0}^{1,1}(\R)}$, a concave $\psi_{c, \mathrm{ccv}} \in \smash{\mathcal{C}_{\geq 0}^{1,1}(\R)}$ with $(-\rho) \leq \psi_{c, \mathrm{ccv}}^{\prime \prime} \leq 0$ a.e., and $\mu \in \R_{\geq 0}$.
   The convex part $\psi_{c, \mathrm{cvx}}$ of the learned $\psi_c$ is necessary to maintain differentiability at $0$.
    Since our $\psi_c$ only takes positive inputs, we do not have this issue and can drop the term $\psi_{c, \mathrm{cvx}}$.
    Further, we relax $\rho$ from $1$ to $\infty$ to fully explore the role of concavity.
    Given that the experimental results are similar, we do not expect that the inclusion of a convex part $\psi_{c, \mathrm{cvx}}$ in \eqref{eq:prp_reg} leads to a significant gain in performance.}
    
    \fixme{For the regularizer  $\mathcal{R}_{\text{SAFI}, k}$, we use the absolute value $\vert \cdot \vert$ instead of non-convex potentials, as proposed in \cite{NeuPouGou2023,NeuAlt2024}. Hence, the subproblem \eqref{eq:set_cvx_prblms} for each SAFI update is convex and the deployed optimization algorithm converges to a minimizer.
    This is stronger than the mere convergence to stationary points of \cite{NeuPouGou2023}.
    Since we learn $\M W$, our $\mathcal{R}_{\text{SAFI}, k}$ generalizes the data-adaptive total-variation model in \cite{KofAltBa2023}.
    Moreover, in contrast to these approaches, we iteratively refine the mask in $\mathcal{R}_{\text{SAFI}, k}$ based on $\V x_{k+1} = T_{\tilde{\M \Lambda}, \M W, \V y} (\V x_k)$.
    This leads to implicit depth, which is a possible explanation for why complex generators \smash{$\tilde{\M \Lambda}$} are not required in our framework.}
    
    \fixme{The majorization-minimization (MM) perspective also shows up in \cite{lefkimmiatis2023learning}, which deploys MM iterations to minimize a spatially adaptive model that is similar to \cite{NeuPouGou2023, NeuAlt2024}.
    To ensure closed-form solutions for the minimization problems \eqref{eq:set_cvx_prblms}, the authors deploy $\vert \cdot \vert^2$ as potentials instead of $\vert \cdot \vert$ in \eqref{eq:RegConvMask}.
    In contrast to the SAFI approach, their masks \smash{$\tilde{\M \Lambda}(\V x_k)$} for the MM iterations are induced completely by the underlying regularizer.}
    
    \section{Preliminaries \label{sec:preliminaries}}
    Throughout this work, $\X \subseteq \R^N$ denotes a closed convex set.
    \paragraph{Concave Functions}
    A function $f \colon \X \to \R$ is said to be \emph{concave} if it satisfies 
        \begin{equation}\label{def:cnv_fun}
            f\bigl(\alpha \V x_1 + (1-\alpha) \V x_2\bigr) \geq \alpha f(\V x_1) + (1-\alpha) f(\V x_2),  \qquad \forall \V x_1, \V x_2 \in \X, \quad \forall \alpha \in [0, 1].
        \end{equation}
        If $\X$ is open and $f \in \mathcal{C}^1(\X)$, then $f$ is concave if and only if its gradient $\V \nabla f$ satisfies
        \begin{equation}\label{eq:mon_gra}
            \bigl \langle \V \nabla f(\V x_1) - \V \nabla f(\V x_2), \V x_1 - \V x_2 \bigr \rangle \leq 0, \qquad \forall \V x_1, \V x_2 \in \X.
        \end{equation}
        In the special case $N=1$, condition \eqref{eq:mon_gra} simply states that the derivative $f^{\prime}$ is non-increasing on $\X$. 
        Another useful property is that any differentiable concave function $f$ is upper-bounded by its first-order Taylor expansion
        \begin{equation}\label{eq:maj_tay}
            f(\V x_1) \leq f(\V x_2) + \langle \V \nabla f(\V x_2), \V x_1 - \V x_2 \rangle, \qquad \forall \V x_1, \V x_2 \in \X.
        \end{equation}
        A function $f \colon \X \to \R$ is \emph{convex} if and only if $(-f)$ is concave. 
        \paragraph{Majorization-Minimization Algorithm}
        For a deeper exposition to MM algorithms, we refer to \cite{HunLan2004,FigBioNow2007,SunBabPal2017}.
        Here, we only collect some basic definitions and the core results.
        For a continuous $f\colon \X \to \R$, we investigate the problem
        \begin{equation}
            \argmin_{\V x \in \X} f(\V x).
        \end{equation}
        The idea behind MM algorithms is to replace $f$ by a sequence of (approximating) majorizations $g(\cdot,\V x_k)$, $\V x_k \in \X$ for which the computation of a (global) minimizer is tractable.
        A function $g \colon \X \times \X \to \R$ is said to be a \emph{majorization} of $f \colon \X \to \R$ if it satisfies
        \begin{enumerate}
            \item[i)] the upper-bound $f(\V x) \leq g(\V x, \V x_k), \quad \forall \V x, \V x_k \in \X$;
            \item[ii)] and the local tight bound $g(\V x_k, \V x_k) = f(\V x_k), \quad \forall \V x_k \in \X.$
        \end{enumerate}
        Next, we introduce the formal MM algorithm together with a convergence result \cite{SunBabPal2017,JacJef2007}.
        \begin{theorem} \label{the:maj_min}
            For a continuous $f \colon \X \to \R$ with majorization $g \colon \X \times \X \to \R$ and a starting point $\V x_1 \in \X$, the MM sequence is given by
            \begin{equation}\label{eq:MMiters}
                \V x_{k+1} \in \argmin_{\V x \in \X} g(\V x, \V x_k),
            \end{equation}
            and the function values $f(\V x_k)$ are non-increasing.
            If $g$ is continuous, $f$ and every $g(\cdot,\V x_k)$ is continuously differentiable, and the sub-level set $\{\V x \in \X: f(\V x) \leq f(\V x_1)\}$ is compact,
            then all accumulation points of $\{\V x_k\}_{k \in \N}$ are critical points of $f$. 
            Moreover, if the set $\X^* = \{\V x : \langle \V \nabla f (\V x), \V z - \V x \rangle \geq 0, \ \forall \V z \in \X\}$ is a singleton or if $\X^*$ is discrete and $\lim_{k \to \infty} \Vert \V x_{k+1} - \V x_k \Vert_2 \to 0$, then the MM iterations $\{\V x_k\}_{k \in \N}$ converge to a critical point of $f$.
        \end{theorem}
        \begin{remark}
            The condition that $\X^*$ is a singleton is met if $f$ is strongly convex, namely if $(f - \tfrac{\sigma}{2}\Vert \cdot \Vert^2_2)$ is convex for some $\sigma \in \R_+$.
            Hence, we get in this setting global convergence guarantees that are similar to those of convex-minimization algorithms.
        \end{remark}

        \paragraph{$\Gamma$-Convergence}
        Here, we recall the basic concepts of $\Gamma$-convergence within our Euclidean framework and refer to \citep{Braides02} for a more detailed exposition.
        A family of functions $\{J_k\}_{k\in\N}$ 
        with $J_k\colon \X \rightarrow [0,\infty]$ is said to $\Gamma$-converge to $J \colon  \X \rightarrow [0,\infty]$ 
        if the following two conditions are fulfilled for every $\V x \in \X$:
        \begin{enumerate}
	       \item[i)] for all $\V x_k  \to  \V x$, it holds that $J(\V x) \leq \liminf_{k \to \infty} J_k(\V x_k)$;
	       \item[ii)] for every $\V x \in \X$, there is a sequence $\{\V x_k\}_{k\in\N}$ with $\V x_k \to \V x$ and $\limsup_{k \to \infty} J_k(\V x_k) \le J(\V x)$.
        \end{enumerate}
        The importance of $\Gamma$-convergence is captured by Theorem~\ref{thm:FundGamma}.
        Recall that a family of functions $J_k \colon \X \to \R$ is equi-coercive if it is bounded from below by a coercive function.
        \begin{theorem}[Theorem of $\Gamma$-convergence \citep{Braides02}]\label{thm:FundGamma}
        Let $\{J_k\}_{k \in \N}$ be an equi-coercive family of functions $J_k \colon \X \to \R$.
        If $J_k$ $\Gamma$-converges to $J$, then it holds that
        \begin{itemize}
            \item[i)] the optimal function values converge $\lim_{k \to \infty} \inf_{\V x \in \X } J_k(\V x) = \inf_{\V x \in \X} J(\V x)$; 
            \item[ii)] all accumulation points of the minimizers of $J_k$ are minimizers of $J$. 
        \end{itemize}
        \end{theorem}
        In particular, if all the $J_k$ and $J$ have unique minimizers, then Theorem~\ref{thm:FundGamma} directly implies convergence of the minimizers of the $J_k$ to the one of $J$.
        
        \section{New Perspectives on Ridge-Based Regularization}
        First, we provide a novel perspective on weakly convex ridge regularizers \cite{GouNeuUns2023} \fixme{through the MMR model.
        Based on this perspective, we then derive our more general SAFI} reconstruction scheme.
        \subsection{Majorization-Minimization Regularization}\label{sec:MajMin}
        \fixme{For the MMR model}, we specify $\mathcal R$ in the generic problem \eqref{eq:dis_inv_pro} as \eqref{eq:prp_reg} \fixme{and choose} $E$ as the squared norm.
        Moreover, we allow for linear constraints by minimizing over a closed convex polytope $\X \subset \R^N$.
        This leads to the problem
        \begin{equation}
            \label{eq:inv_spc}
            \argmin_{\V x \in \X} f(\V x) \coloneqq \Big( \frac{1}{2} \norm{\M H \V x - \V y }_2^2 + \lambda \sum_{c=1}^{ N_\mathrm{C}} \bigl \langle \mathbf{1}_{\fixme{N}}, \boldsymbol \psi_c  \big( \M B_{c}  \vert \M W_{c} \V x \vert\bigr)\bigr\rangle \Big).
        \end{equation}
        First, we establish the existence of minimizers for \eqref{eq:inv_spc}.
        \begin{theorem}\label{thm:existence}
            Let $\psi_c\colon \R \rightarrow \R_{\geq 0}$, $c=1,\ldots,N_C$, be continuous and piecewise-polynomial functions with finitely many pieces, and let $\X \subset \R^N$ be a closed convex polytope.
            Then, problem \eqref{eq:inv_spc} admits a minimizer. 
        \end{theorem}
    \begin{proof}
    Each $\psi_c$ partitions $\R$ into finitely many closed\footnote{Such a partition with closed interval exists because $\psi_c$ is continuous.} intervals $(I_c^{m})_{m=1}^{L_c}$ on which it is a polynomial.
    Hence, if we denote the $n$-th row of $\M B_c$ by $\M B_{c,n}$, each $\boldsymbol \psi_c(\M B_{c,n} \vert \M W_c \cdot\vert)$ partitions $\X$ into $L_c$ closed unions of polytopes $\Omega_{c,n}^{m}=\{\V x \in \R^N : \M B_{c,n} \vert \M W_c \V x \vert \in I_c^{m}\}$.
    Based on these, we can further partition $\X$ into finitely many closed polytopes, each of which is contained in one of the  \smash{$\cap_{c,n=1}^{N_C,N} \Omega_{c,n}^{m_{c,n}}$}, where $m_{c,n} \in \{1,\ldots,L_c\}$, and on which all the $\M B_{c,n} \vert \M W_c \cdot \vert$ are linear.
    The infimum in \eqref{eq:inv_spc} is the infimum of $f$ on (at least) one of these polytopes, say $P$.
    
    Now, we pick a minimizing sequence $(\V x_k)_{k\in \N}\subset P$.
    Due to the coercivity of $\|\cdot\|_2^2$, we get that the sequence $(\M H \V x_k)_{k\in\N}$ remains bounded.
    By construction, there exist diagonal matrices $\M D_{c} \in \R^{N,N}$ such that $\M B_{c,n} \vert \M W_{c} \V x \vert = \M B_{c,n} \M D_c \M W_{c} \V x$ for every $n=1,\dots,N$ and $\V x\in P$.
    Let $\M M$ be the matrix which is the vertical concatenation of $\M H$ and all the $\M B_{c,n} \M D_c \M W_{c}$ with $c$ and $n$ such that $(\M B_{c,n} \M D_c \M W_{c} \V x_k)_{k \in \N}$ remains bounded.
    Since the sequence $(\M M \V x_k)_{k\inN}$ is bounded, we can extract a convergent subsequence with limit $\V u \in \text{ran}(\M M)$.
    The associated set
    \begin{equation}
    Q=\{\V x\in\R^N \colon \M M\V x = \V u\} = \{\M M^\dagger \V u\} + \ker(\M M)
    \end{equation}
    is a closed polytope.
    It holds that
    \begin{align}
    \mathrm{dist}(\V x_k, Q) &= \mathrm{dist}\bigl(\M M^\dagger \M M \V x_k + \mathrm P_{\ker(\M M)}(\V x_k), Q\bigr) \nonumber\\
    & \leq \mathrm{dist}(\M M^\dagger \M M \V x_k, \M M^\dagger \V u) \to 0
    \end{align}
    as $k\to +\infty$ and, thus, that $\mathrm{dist}(P, Q)=0$.
    The distance of $P$ and $Q$ is 0 if and only if $P\cap Q \neq \emptyset$ \cite[Thm.~1]{W1968}.
    For the $\M B_{c,n} \M D_c \M W_{c}$ that were not added to $\M M$, it holds that $\M B_{c,n} \vert \M W_{c} \V x_k \vert \to \infty$.
    Hence, the interval \smash{$I_c^{m_{c,n}}$} has to be unbounded.
    Since $\psi_c$ is a nonnegative polynomial on it, $\boldsymbol \psi_c(\M B_{c,n} \vert \M W_{c} \cdot \vert)$ has to be constant\footnote{A non-constant polynomial cannot have a finite limit at $\pm\infty$.} on $P$ and $\boldsymbol \psi_c(\M B_{c,n} \vert \M W_{c} \V x_k \vert) = \boldsymbol \psi_c(\M B_{c,n} \vert \M W_{c} \V x \vert)$ for every $\V x\in P\cap Q$.
    Hence, any $\V x\in P\cap Q$ is a minimizer for \eqref{eq:inv_spc}.
    \end{proof}
    \begin{remark}
        A crucial ingredient for our proof is the architecture \eqref{eq:prp_reg} with $\vert \cdot \vert$ as the inner nonlinearity.
        In general, it is much harder to guarantee the existence of minimizers for piecewise-polynomial functions \cite{Pha2023}.
    \end{remark}
        The $f$ in \eqref{eq:inv_spc} is not necessarily convex.
        Hence, one should not attempt to solve \eqref{eq:inv_spc} using conventional convex-optimization algorithms.
        Instead, one can use the majorization-minimization (MM) algorithm defined in \eqref{eq:MMiters} to search for stationary points.
        When $f$ is convex, this algorithm converges to a minimizer.
        To apply the MM algorithm, we first show that the concavity of the $\psi_c \in \mathcal{C}_{\geq0}^{1,1}(\R)$ implies the concavity of $g_c(\V x) = \langle \mathbf{1}_{\fixme{N}}, \boldsymbol \psi_c (\M B_c \V x) \rangle$.
        Based on this property, we then construct a majorization of \fixme{$\mathcal{R}_{\text{MMR}}$}.
        \begin{lemma}\label{pro:cnv_g}
            If $\psi_c \in \smash{C_{\geq0}^{1,1}(\R)}$, $\ c = 1, \ldots, N_C$, then $g_c$ is differentiable with $\V \nabla g_c(\V x) = \M B_c^{\top} \boldsymbol \psi_c^{\prime}(\M B \V x)$.
            Moreover, if $\psi_c$ is also concave, then $g_c$ is concave as well.
        \end{lemma}
        \begin{proof}
            We have that $g_c(\V x) = h(\boldsymbol \psi_c(\M B_c \V x))$ with $h(\V x) = \langle \V 1_N, \V x \rangle$.
            Hence, the Jacobian $\V J_g$ is given through the chain rule as 
            \begin{equation}
                    \V J_{g_c} (\V x) = \V J_{h \circ \boldsymbol \psi_c \circ \M B_c}(\V x) = \V J_{h} (\boldsymbol \psi_c(\M B \V x)) \V J_{\psi_c} (\M B \V x) \M B = \V 1_N^{\top} \mathbf{diag}(\boldsymbol \psi_c^{\prime}(\M B_c \V x)) \M B_c =  \boldsymbol \psi_c^{\prime}( \M B_c \V x)^{\top} \M B_c,
                \end{equation}
            where $\mathbf{diag} \colon \R^N \to \R^{N, N}$ returns a diagonal matrix whose diagonal elements are the input vector.
            As $ \V \nabla g (\V x) = \V J_{g} (\V x)^{\top}$, the first claim readily follows. 
            Further, it holds for $\V x_1, \V x_2 \in \R^N$ that
            \begin{align}
                \label{eq:prf_cnv}
                &\bigl \langle \V \nabla g_c(\V x_1) - \V \nabla g_c(\V x_2), \V x_1 - \V x_2 \bigr \rangle = \bigl \langle \M B_c^{\top} \boldsymbol\psi_c^{\prime}(\M B_c \V x_1) -   \M B_c^{\top} \boldsymbol \psi_c^{\prime}(\M B_c  \V x_2), \V x_1 - \V x_2 \bigr \rangle \notag\\
                = &\bigl \langle \M B_c^{\top}(\boldsymbol \psi_c^{\prime}(\M B_c  \V x_1) -  \boldsymbol \psi_c^{\prime}(\M B_c \V x_2)), \V x_1 - \V x_2 \bigr \rangle = \bigl\langle \boldsymbol \psi_c^{\prime}(\M B_c  \V x_1) -  \boldsymbol \psi_c^{\prime}(\M B_c \V x_2), \M B_c (\V x_1 - \V x_2) \bigr\rangle \notag\\
                = &\langle \boldsymbol \psi_c^{\prime}(\M B_c  \V x_1) -  \boldsymbol \psi_c^{\prime}(\M B_c \V x_2), \M B_c \V x_1 - \M B_c \V x_2 \rangle \leq 0, 
            \end{align}
            where the inequality stems from the concavity of $\psi_c$.
            By \eqref{eq:mon_gra}, the $g_c$ are concave and the proof is complete. 
        \end{proof}
        Now, we majorize $g_c$ using its first-order Taylor expansion, see \eqref{eq:maj_tay}, and get that
        \begin{equation}
            \label{eq:maj_g}
            \bigl \langle \mathbf{1}_{\fixme{N}}, \boldsymbol \psi_c (\M B_c \V x) \bigr\rangle \leq \bigl \langle \mathbf{1}_{N}, \boldsymbol \psi_c (\M B_c \V x_k) \bigr\rangle + \bigl \langle \M B_c^{\top} \boldsymbol \psi_c^{\prime}(\M B \V x_k), \V x - \V x_k \bigr\rangle, \qquad \forall \V x_k \in \R^N.
        \end{equation}
        With the change of variables $\V x \mapsto \vert \M W_c \V x\vert$ and by summing over all $c$, we then get for any $\V x_k \in \R^N$ that 
        \begin{align}\label{eq:maj_reg}
            \fixme{\mathcal R_{\text{MMR}}} (\V x) = \sum_{c=1}^{ N_\mathrm{C}} \bigl \langle \mathbf{1}_{\fixme{N}}, \boldsymbol \psi_c (\M B_c\vert \M W_c \V x\vert) \bigr\rangle \leq \sum_{c=1}^{ N_\mathrm{C}} \bigl \langle \mathbf{1}_{\fixme{N}}, \boldsymbol \psi_c (\M B_c\vert \M W_c \V x_k\vert) \bigr\rangle + \sum_{c=1}^{ N_\mathrm{C}} \bigl \langle \boldsymbol \Lambda_c(\V x_k),\vert \M W_c \V x\vert - \vert \M W_c \V x_k\vert \bigr\rangle,
        \end{align} 
        where $\boldsymbol \Lambda_c(\V x) = \M B_c^{\top} \boldsymbol \psi_c^{\prime} (\M B_c\vert \M W_c \V x\vert)$.
        Regarding the notation from Section~\ref{sec:preliminaries}, we choose 
        \begin{align}
            g(\V x, \V x_k) = \frac{1}{2} \norm{\M H \V x - \V y}_2^2 + \lambda \sum_{c=1}^{ N_\mathrm{C}} \bigl \langle \mathbf{1}_{N}, \boldsymbol \psi_c (\M B_c\vert \M W_c \V x_k\vert) \bigr\rangle + \lambda \sum_{c=1}^{ N_\mathrm{C}} \bigl \langle \boldsymbol \Lambda_c(\V x_k),\vert \M W_c \V x \vert - \vert \M W_c \V x_k\vert \bigr\rangle. 
        \end{align}
        It is easy to verify that the chosen $g(\V x, \V x_k)$ is a valid majorization of $f$ in \eqref{eq:inv_spc}.
        Therefore, to compute a stationary point of $f$ based on \eqref{eq:MMiters}, we have to compute the estimates
        \begin{equation}
        \label{eq:set_cvx_prblms2}
        \V x_{k+1} \in \argmin_{\V x \in \X} \Big(\frac12 \norm{\M H \V x - \V y }_2^2  + \lambda \sum_{c=1}^{ N_\mathrm{C}} \bigl \langle \boldsymbol \Lambda_c(\V x_k),\vert \M W_c \V x\vert\bigr\rangle\Big).
        \end{equation}   
        As $\psi_c^{\prime} \geq 0$, these majorizations of the original problem can be interpreted as spatially reweighted $\ell_1$-analysis regularization, where the strength of the convex summands $\Vert \M W_c \V x_k \Vert_1$ is reweighted by $\boldsymbol \Lambda_c(\V x_k)$.
        Accordingly, we rewrite the convex problem of \eqref{eq:set_cvx_prblms2} in a more compact form as 
        \begin{equation}
            \label{prb:gen_lasso}
            \V x_{k+1} \in \argmin_{\V x \in \X} \Big(\frac12 \norm{\M H \V x - \V y }_2^2  + \lambda \norm{\M L_k \V x}_1\Big) \qquad \text{with } \M L_k = \bigl[\mathbf{diag}(\boldsymbol \Lambda_c(\V x_k)) \M W_{c}\bigr]_{c=1}^{N_C}.
        \end{equation}
        In Algorithm~\ref{alg:FBS1}, we provide an iterative procedure based on FBS \cite{beck2009fast,chambolle2015convergence} to compute \eqref{prb:gen_lasso}.
        To this end, we choose $\frac12 \norm{\M H \cdot - \V y }_2^2$ as the differentiable part of the objective and $\lambda \norm{\M L_k \cdot}_1$ for the non-differentiable one. 
        The most time-consuming part in Algorithm~\ref{alg:FBS1} for generic $\M L$ is the evaluation of the proximal operator $\prox_{\alpha \lambda \norm{\M L \cdot}_1}$ defined as
        \begin{equation}
            \label{eq:prox_g}
            \prox_{\alpha \lambda \norm{\M L \cdot}_1} (\V z)= \argmin\limits_{\V w \in \X} \Big(\frac{1}{2} \norm{\V w - \V z}_2^2 + \alpha \lambda \norm{\M L \V w}_1\Big).
        \end{equation}
        For computational purposes, it is better to consider the dual problem of \eqref{eq:prox_g}, which we derive as in \cite{BecTeb2009}.
        \begin{proposition}
            Let $ \mathrm{Proj}_{\mathcal{X}}\colon \R^N \to \R^N$ denote the orthogonal projection onto $\X$.
            If $\hat{\V u}$ solves the problem 
            \begin{equation}\label{eq:dual_prox}
            \argmin\limits_{\V u \in \R^{N_C \cdot N}} \Big(\frac{1}{2} \Vert \M L^{\top} \V u - \V z\Vert_2^2 - \frac{1}{2} \bigl\Vert \mathrm{Proj}_{\mathcal{X}}\{\M L^{\top} \V u - \V z\} - (\M L^{\top} \V u - \V z) \bigr\Vert_2^2\Big) \quad \text{subject to } \norm{\hat{\V u}}_{\infty} \leq \alpha \lambda,
            \end{equation}
            then $\mathrm{Proj}_{\mathcal{X}}\{\V z - \M L^{\top} \hat{\V u}\}$ equals \eqref{eq:prox_g}.
        \end{proposition}
        \begin{proof}
            By duality, we have that 
            \begin{equation}
                \alpha \lambda \norm{\M L \V w}_1 = \max_{\V u}\bigl\{\V u^{\top} (\M L \V w) : \norm{\V u}_{\infty} \leq \alpha \lambda\bigr\}.
            \end{equation}
            Plugging this into \eqref{eq:prox_g} leads to
            \begin{align}
                &\min_{\V w \in \X} \max_{u}  \Big(\frac{1}{2} \norm{\V w}_2^2 + \frac{1}{2} \norm{\V z}^2_2 + \V w^{\top} (\M L^\top \V u  -\V z)\Big) \quad \text{subject to } \norm{\V u}_{\infty} \leq \alpha \lambda \notag\\
                =&\min_{\V w \in \X} \max_{u}  
                \Big(\frac{1}{2} \norm{\V w - (\V z -\M L^\top \V u  )}^2_2 - \frac{1}{2} \norm{\V z - \M L^\top \V u}^2_2 + \frac{1}{2} \norm{\V z}^2_2\fixme{\Big)} \quad \text{subject to } \norm{\V u}_{\infty} \leq \alpha \lambda .
            \end{align}
            Now, we can swap the $\min$ and $\max$ because the objective is convex in $\V w$ and concave in $\V u$ \cite[Cor.\ 37.3.2]{Rockafellar1997}. 
            Then, we directly get that $\V w = \mathrm{Proj}_{\mathcal{X}}\{\V z - \M L^{\top} \V u\}$ is optimal for the inner minimization.
            By removing the constant term $\frac{1}{2} \norm{\V z}^2_2$ and a change of sign, we obtain \eqref{eq:dual_prox}. 
        \end{proof}
        To solve \eqref{eq:dual_prox}, we apply once again FBS with the objective as the differentiable part and the constraints for the non-differentiable one, see Algorithm~\ref{alg:2}.
        Note that the subtrahend in \eqref{eq:dual_prox} is the concatenation of a Moreau envelope with the affine map $\V u \mapsto (\M L^{\top} \V u - \V z)$.
        Hence, its gradient reads $(\M L (\M L^{\top} \V u - \V z)  - \M L \mathrm{Proj}_{\mathcal{X}}\{\M L^{\top} \V u - \V z\})$, and the overall gradient of the objective in \eqref{eq:dual_prox} with respect to $\V u$ is 
        \begin{equation}
            \M L (\V z - \M L^{\top} \V u) - \M L \big((\V z - \M L^{\top} \V u) - \mathrm{Proj}_{\mathcal{X}}\{\V z - \M L^{\top} \V u\}\big) = \M L \mathrm{Proj}_{\mathcal{X}}\{\V z - \M L^{\top} \V u\}.
        \end{equation}
        Our last ingredient is the saturating function $\textbf{clip}_{[\kappa_1, \kappa_2]}\colon \R^N \to \R^N$, which is defined component-wise as
        \begin{equation}
            [\textbf{clip}_{[\kappa_1, \kappa_2]}(\V a)]_k = \mathrm{clip}_{[\kappa_1, \kappa_2]}(a_k) = \begin{cases}
            \kappa_1, & a_k < \kappa_1 \\
            a_k, & \kappa_1 \leq a_k \leq \kappa_2 \\
            \kappa_2, & {a_k} > \kappa_2.
        \end{cases}
        \end{equation}
        
        Our proposed \fixme{MMR} scheme is summarized in Algorithm~\ref{alg:MajMin}.
        It deploys the FBS (Algorithm~\ref{alg:FBS1}) to solve the majorization-minimization problems \eqref{prb:gen_lasso}.
        If $\M H = \text {Id}$, Algorithm~\ref{alg:FBS1} can be terminated after one step.
        The involved operator $\mathbf{Prox}_{\alpha \lambda \Vert \M  L_k \cdot \Vert_1}$ is computed using again the FBS (Algorithm~\ref{alg:2}).
        \fixme{For both algorithms, our choice of $\{t_k\}_{k \in \N}$ ensures the convergence of the iterates \cite{chambolle2015convergence}.}
        Under the assumption of infinite precision in the sub-routines, Algorithm~\ref{alg:MajMin} finds indeed a critical point of $f$. 
        
        \begin{theorem}\label{thm:convergnece}
            Assume that the estimates \eqref{prb:gen_lasso} are obtained exactly within Algorithm~\ref{alg:MajMin}.
            Then, $\{f(\V x_k)\}_{k \in \N}$ is non-increasing.
            If $\M H$ is invertible,
            then $f$ is coercive and all accumulation points of $\{\V x_k\}_{k \in \N}$ are in the set of critical points
            \begin{equation}
                \X^* = \Bigl\{\V x_1 \in \X: \langle \M H^{\top}(\M H \V x_1 - \V y), \V x_2 - \V x_1 \rangle + \lambda \sum_{c=1}^{N_C} \bigl \langle \boldsymbol \Lambda_c(\V x_1), \vert \M W_c \V x_2 \vert  - \vert \M W_c \V x_1 \vert \bigr\rangle \geq 0,\ \forall \V x_2 \in \X \Bigr\}.
            \end{equation} 
            Moreover, if $\X^*$ is a singleton or if $\X^*$ is discrete and $\lim_{k \to \infty} \Vert \V x_{k+1} - \V x_k \Vert_2 \to 0$, then the MM iterates $\{\V x_k\}_{k \in \N}$ converge to a critical point of $f$.
        \end{theorem}
        \begin{proof}
            First, we introduce the auxiliary variable $\V z \in \R_{\geq 0}^{N_C \cdot \N}$ with grouped components $\V z_c = \vert \M W_{c} \V x \vert \in \R^N$ in \eqref{eq:inv_spc} and investigate the equivalent problem
            \begin{equation}
            \label{eq:inv_spc_lifted}
            \argmin_{\V x \in \X} \min_{\V z \in \R_{\geq 0}^{N_C \cdot \N}} \tilde f(\V x, \V z) \coloneqq \frac{1}{2} \norm{\M H \V x - \V y }_2^2 + \lambda \sum_{c=1}^{ N_\mathrm{C}} \bigl \langle \mathbf{1}_{\fixme{N}}, \boldsymbol \psi_c  \big( \M B_{c} \V z_c \bigr)\bigr\rangle \quad \text{subject to } \V z_c = \vert \M W_{c} \V x \vert.
            \end{equation}
            Then, the majorizations will take the form
            \begin{align}
            \tilde g\bigl((\V x,\V z), (\V x_k, \V z_k)\bigr) = \frac{1}{2} \norm{\M H \V x -\V y}_2^2 + \lambda \sum_{c=1}^{ N_\mathrm{C}} \bigl \langle \mathbf{1}_{\fixme{N}}, \boldsymbol \psi_c (\M B_c \V z_{k,c}) \bigr\rangle + \lambda \sum_{c=1}^{ N_\mathrm{C}} \bigl \langle \M B_c^{\top} \boldsymbol \psi_c^{\prime} (\M B_c\V z_{k,c}), \V z  -\V z_{k,c} \bigr\rangle,
            \end{align}
            and their minimization subject to $\V z_c = \vert \M W_{c} \V x \vert$ leads indeed to \eqref{eq:set_cvx_prblms}.
            Observe that $\tilde g$ are continuous.
            Further, both $\tilde f$ and the $\tilde g(\cdot, (\V x_k, \V z_k))$ are differentiable.
            Hence, we can apply Theorem~\ref{the:maj_min} and the claim follows.
        \end{proof}
        \begin{algorithm}[t]
        \caption{FBS for solving \eqref{prb:gen_lasso}}
        \label{alg:FBS1}
        \begin{algorithmic}[1]
        \State \textbf{Input}: filter matrix $\M L$, previous minimizer $\V x_1$, current iteration $k_{\mathrm{out}}$\;
        \State \textbf{Parameters}: maximal iteration number $K_{\mathrm{FBS}}$, dynamic tolerance $\epsilon_{\mathrm{FBS}} = f_{\epsilon,\mathrm{FBS}} (k_{\mathrm{out}})> 0$\;
        \State \textbf{Initialize}: $t_1 = 1$, $\alpha = 1 / \Vert \M H \Vert_2^2$, $\tilde{\V x}_1 = \V x_1$\;
        \For{$k = 1$ \textbf{to} $K_{\mathrm{FBS}}$}
        \State $\V x_{k+1}={\mathbf{Prox}}_{\alpha \lambda \Vert \M  L \cdot \Vert_1}(\tilde{\V x}_k - \alpha \M {H}^{\top} (\M {H} \tilde{\V x}_k - \V y), k_{\mathrm{out}}, k)$\;
        \State $t_{k+1} = \fixme{(k+5) / 3}$\;
        \State $\tilde{\V x}_{k+1} = \V x_{k+1} + \frac{t_k - 1}{t_{k+1}}  (\V x_{k+1} - \V x_k)$;
        \If{$\norm{\V x_{k+1} - \V x_k}_2 < \epsilon_{\mathrm{FBS}} \norm{\V x_k}_2$}
        \State \textbf{break}
        \EndIf
        \EndFor
        \State \textbf{return} $\V x_{k+1}$ 
        \end{algorithmic}
        \end{algorithm}
        
        \begin{algorithm}[t]
        \caption{Computation of $\prox_{\gamma \Vert \M  L \cdot \Vert_1}$ based on the dual \eqref{eq:dual_prox} using FBS}
        \label{alg:2}
        \begin{algorithmic}[1]
        \State \textbf{Input}: vector $\V z \in \R^{N}$, current iteration $k_{\mathrm{out}}$, current iteration $k_{\mathrm{FBS}}$\;
        \State \textbf{Parameters}: maximal iteration number $K_{\mathrm{prox}}$, dynamic tolerance $\epsilon_{\mathrm{prox}} = f_{\epsilon,{\mathrm{prox}}}(k_{\mathrm{out}}, k_{\mathrm{FBS}}) > 0$\;
        \State \textbf{Initialize}: $\V u_1 = \M L \V z$, $\V v_1 = \M L \V z$, $\V x_1 = \mathrm{Proj}_{\mathcal{X}}\{\V z - \M L^{\top} \V u_1\}$, $t_1 = 1$, $\alpha = 1 / \Vert \M L\Vert_2^2$\; 
        \For{$k = 1$ \textbf{to} $K_{\mathrm{prox}}$}
        \State $\V u_{k+1} = \textbf{clip}_{[-\gamma, \gamma]}(\V v_k - \alpha \M L \mathrm{Proj}_{\mathcal{X}}\{\M L^{\top} \V v_k - \V z\})$\;
        \State $t_{k+1} = \fixme{(k+5)/3}$\;
        \State $\V v_{k+1} = \V u_{k+1} + \frac{t_k - 1}{t_{k+1}}  (\V u_{k+1} - \V u_k)$\;
        \State $\V x_{k+1} = \mathrm{Proj}_{\mathcal{X}}\{\V z - \M L^{\top} \V u_{k+1}\}$
        \If{$\norm{\V x_{k+1} - \V x_k}_2 < \epsilon_{\mathrm{prox}} \norm{\V x_k}_2$}
        \State \textbf{break}
        \EndIf
        \EndFor
        \State \textbf{return} $\V x_{k+1}$
        \end{algorithmic}
        \end{algorithm}

        \begin{algorithm}[t]
        \caption{\fixme{MMR scheme} for \eqref{eq:inv_spc}}
        \label{alg:MajMin}
        \begin{algorithmic}[1]
        \State \textbf{Parameters}: maximal iteration number $K_{\mathrm{out}}$, tolerance $\epsilon_{\mathrm{out}} > 0$\;
        \State \textbf{Initialize}: ${\V x_1  = \V 0, \M L_1 = [\M W_c]_{c=1}^{N_C}}$\; 
        \For{$k = 1$ \textbf{to} $K_{\mathrm{out}}$}
        \State $\V x_{k+1} = \mathrm{\textbf{FBS}}(\M L_k, \V x_k, k)$\;
        \State Compute $\M L_{k+1} = [\mathbf{diag} (\M \Lambda_c(\V x_{k+1})) \M W_{c}]_ {c=1}^{N_C}$\;
        \If{$ \norm{\V x_{k+1} - \V x_k}_2 < \epsilon_{\mathrm{out}} \norm{\V x_k}_2$}
        \State \textbf{break}
        \EndIf
        \EndFor
        \State \textbf{return} $\V x_{k+1}$ 
        \end{algorithmic}
        \end{algorithm}
        
        \subsection{Solution-Adaptive Fixed-Point Iterations}\label{sec:SolDrivAdapt}
        For the \fixme{MMR model} with \eqref{eq:set_cvx_prblms}, the mask generator ${\M \Lambda} \colon\R^N \to \smash{(\R_{\geq 0}^N)}^{N_C}$ allows for a successive spatial adaption of the regularization strength.
        So far, the architecture of each $\M \Lambda_c \colon \R^N \to \R_{\geq0}^N$ is motivated by the \fixme{MMR} perspective.
        One might wonder if a more generic $\smash{\tilde{\M \Lambda} \colon \R^N \to ([0, 1]^N)^{N_C}}$ leads to improvements. \fixme{This leads to the SAFI scheme based on \eqref{eq:RegConvMask}.}
        As the masks generated by \smash{$\tilde{\M \Lambda}_c$} are nonnegative, the resulting \fixme{SAFI} updates
        \begin{equation}\label{eq:set_cvx_prblms_generic}
        \V x_{k+1} \in \argmin_{\V x \in \X} J(\V x, \V x_k) \coloneqq \frac12 \norm{\M H \V x - \V y }_2^2  + \lambda \sum_{c=1}^{ N_\mathrm{C}}  \bigl \langle \smash{\tilde{\M \Lambda}_c}(\V x_k),\vert \M W_c \V x\vert\bigr\rangle
        \end{equation}
        are \fixme{minimizers of} convex \fixme{problems}.
        Moreover, if $\M H$ is invertible, then \eqref{eq:set_cvx_prblms_generic} is a singleton.
        Hence, this case gives rise to an update operator $T_{\M  \Lambda, \M W, \V y}\colon \X \to \X$.
        For the \fixme{MMR} framework in Section~\ref{sec:MajMin} with $\M \Lambda_c (\V x) = \M B_c^{\top} {\boldsymbol \psi}_c^{\prime}(\M B_c \abs{\M W_{c} \V x})$, we are guaranteed that the fixed-point iterates
        \begin{equation}\label{eq:FixedPointIters}
            \V x_{k+1} = T_{\M \Lambda, \M W, \V y} (\V x_k)
        \end{equation}
        are convergent.
        Moreover, the resulting fixed point is a \fixme{critical point} of problem \eqref{eq:inv_spc}.
        
        If one is only interested in obtaining convergence of the iterates \eqref{eq:FixedPointIters}, this choice is overly constraining.
        For example, convergence can be guaranteed whenever $T_{\M \Lambda, \M W, \V y}$ is nonexpansive.
        Any fixed point $\V x^*$ of \eqref{eq:FixedPointIters} is a critical point of \eqref{eq:set_cvx_prblms_generic} with $\V x_k = \V x^*$,
        and we cannot \emph{improve} $\V x^*$ by updating the $\smash{\tilde{\M \Lambda}_c}$ anymore.
        In contrast to \cite{NeuPouGou2023}, the spatial adaptivity of the SAFI is driven by every estimate \eqref{eq:set_cvx_prblms_generic}, and not only by an initial reconstruction based on the data $\V y$.
        For the special case of total-variation regularization, this was therefore coined as solution-driven adaptivity instead of data-driven adaptivity \cite{LenLelBec2014,LenBer2015}.
        In general, the updates \eqref{eq:FixedPointIters} are unrelated to the \fixme{critical points} of the non-convex minimization problem
        \begin{equation}
        \argmin_{\V x \in \X} \Big(\frac12 \norm{\M H \V x - \V y }_2^2  + \lambda \sum_{c=1}^{ N_\mathrm{C}}  \bigl \langle \smash{\tilde{\M \Lambda}_c}(\V x),\vert \M W_c \V x\vert\bigr\rangle \Big),
        \end{equation}
        where one also minimizes over the input of $\smash{\tilde{\M \Lambda}_c}$.
        
        Following the discussed ideas, we proceed as outlined in Figure~\ref{fig:freu_box} and Section~\ref{sec:Intro}, and replace $\M \Lambda_c (\V x)$ by the richer architecture $\smash{\tilde{\M \Lambda}_c}(\V x) = \boldsymbol \phi_{3, c}(\hat{\M B}_{c} {\boldsymbol \phi}_{2} (\tilde{\M B} {\boldsymbol \phi}_1 (\tilde{\M W} \V x)))$.
        As observed in \cite{NeuPouGou2023}, one expects that $\smash{\tilde{\M \Lambda}_c}$ dampens the response of the $\M W_c$ to structure and leaves it unchanged for noise or artifacts.
        Under some conditions, we can show that $T_{\tilde{\M \Lambda}, \M W, \V y}$ admits indeed at least one fixed point.
        Hence, the definition of reconstructions as fixed points of the operator $T_{\tilde{\M \Lambda}, \M W, \V y}$ makes sense.
        \begin{theorem}\label{thm:ExistFix}
            Let $\M H$ be invertible and let $\sigma_{\min}$ denote its smallest singular value.
            Then, $T_{\tilde{\M \Lambda}, \M W, \V y}\colon \X \to \X$ maps $\mathcal{X}$ into a ball centered at $\V 0$ with radius $2 \norm{\V y }_2/\sigma_{\min}$.
            Further, $T_{\tilde{\M \Lambda}, \M W, \V y}$ admits a fixed point.
        \end{theorem}
        \begin{proof}
           First, we investigate the range of $T_{\tilde{\M \Lambda}, \M W, \V y}$.
            By definition of $T_{\tilde{\M \Lambda}, \M W, \V y}$, it holds for any $\V x \in \X$ that
            \begin{equation}
                 \frac12 \bigl \Vert \M H T_{\tilde{\M \Lambda}, \M W, \V y}(\V x) - \V y \bigr \Vert_2^2 \leq J(T_{\tilde{\M \Lambda}, \M W, \V y}(\V x), \V x) \leq J(\V 0, \V x) = \frac12 \norm{\V y }_2^2.
            \end{equation}
            From this, we conclude that 
            \begin{equation}
                \bigl\Vert T_{\tilde{\M \Lambda}, \M W, \V y}(\V x)\bigr\Vert_2 \leq \frac{1}{\sigma_{\min}} \bigl \Vert \M H T_{\tilde{\M \Lambda}, \M W, \V y}(\V x)\bigr \Vert_2 \leq 2 \frac{\norm{\V y }_2}{\sigma_{\min}}.
            \end{equation}
            
            For the second part, we want to apply Brouwer's fixed-point theorem.
            To this end, we additionally need to prove that $T_{\tilde{\M \Lambda}, \M W, \V y}$ is continuous.
            Due to Theorem~\ref{thm:FundGamma}, it suffices to check equi-coercivity and the conditions for $\Gamma$-convergence of the family $J(\cdot, \V x)$ parameterized by $\V x \in \X$.
            First, note that it holds for any $\V z \in \X$ that
            \begin{equation}
                \frac{\sigma_{\min}^2}{4} \Vert \V z \Vert_2^2 \leq  \frac{1}{4} \Vert \M H \V z \Vert_2^2 \leq \frac{1}{2} \Vert \M H \V z - \V y \Vert_2^2 +  \Vert \V y \Vert_2^2 \leq J(\V z, \V x) + \Vert \V y \Vert_2^2,
            \end{equation}
            which implies equi-coercivity.
            Now, let $\V x_k \to \V x^*$ and $\V z_k \to \V z^*$.
            By the triangle inequality, we get that 
            \begin{equation}
                \vert J(\V z^*, \V x^*) - J(\V z_k, \V x_k) \vert \leq \vert J(\V z^*, \V x^*) - J(\V z^*, \V x_k)\vert + \vert J(\V z^*, \V x_k) - J(\V z_k, \V x_k)\vert.
            \end{equation}
            To obtain the $\liminf$ inequality, it suffices to prove that the first two terms converge to $0$ as $k \to \infty$.
            For the first one, we have that
            \begin{equation}
                \vert J(\V z^*, \V x^*) - J(\V z^*, \V x_k)\vert \leq  \lambda \sum_{c=1}^{ N_\mathrm{C}}  \bigl \langle \vert \tilde{\boldsymbol \Lambda}_c(\V x^*) - \tilde{\boldsymbol \Lambda}_c(\V x_k) \vert,\vert \M W_c \V z^*\vert\bigr\rangle \to 0
            \end{equation}
            because $\M \Lambda_c$ is continuous.
            For the second one, we have that
            \begin{align}\label{eq:LimSup}
                \vert J(\V z^*, \V x_k) - J(\V z_k, \V x_k)\vert & \leq  \frac{1}{2} \Bigr \vert \Vert \M H \V z^* - \V y \Vert_2^2 - \Vert \M H \V z_k - \V y \Vert_2^2 \Bigr \vert + \lambda \sum_{c=1}^{ N_\mathrm{C}}  \bigl \langle \vert \tilde{\boldsymbol \Lambda}_c(\V x_k) \vert,\bigl \vert\vert \M W_c \V z^*\vert - \vert \M W_c \V z_k\vert \bigr\vert\bigr\rangle \notag\\
                & \leq \frac{1}{2} \Bigr \vert \Vert \M H \V z^* - \V y \Vert_2^2 - \Vert \M H \V z_k - \V y \Vert_2^2 \Bigr \vert + \lambda \sum_{c=1}^{ N_\mathrm{C}}  \bigl \langle \V 1_N, \vert \M W_c \V z^* - \M W_c \V z_k\vert\bigr\rangle.
            \end{align}
            Again, we conclude that this quantity converges to zero.
            Hence, we have established the $\liminf$ inequality.
            For the $\limsup$ inequality, we use the constant recovery sequence $\V x_k = \V x^*$, for which the claim follows as in \eqref{eq:LimSup}.
            In summary, this implies that $T_{\tilde{\M \Lambda}, \M W, \V y}$ is continuous and that a fixed point exists.
        \end{proof}
        \begin{remark}
            Based on a quasi-variational inequality perspective, the authors of \cite{LenLelBec2014} prove the uniqueness of fixed points for certain problems of the form \eqref{eq:FixedPointIters}.
            Unfortunately, their assumptions are hard to verify in practice for $\tilde{\M \Lambda}_c$.
            Hence, we do not pursue this direction further and only provide a proof of existence.
        \end{remark}
        For finding a fixed point \fixme{of the SAFI operator $T_{\tilde{\M \Lambda}, \M W, \V y}$}, we propose to use the fixed-point iterations \eqref{eq:FixedPointIters} detailed in Algorithm~\ref{alg:FixIters}.
        Unfortunately, a proof of convergence for these iterations is highly nontrivial.
        In practice, we encourage this property \fixme{by using a random number of iterations} for the training of the model, as detailed in Section~\ref{sec:Training}.
        \begin{algorithm}[t]
        \caption{\fixme{SAFI} scheme for \eqref{eq:FixedPointIters}}
        \label{alg:FixIters}
        \begin{algorithmic}[1]
        \State \textbf{Parameters}: maximal iteration number $K_{\mathrm{out}}$, tolerance $\epsilon_{\mathrm{out}} > 0$\;
        \State \textbf{Initialize}: ${\V x_1 = \V 0, \M L_1 = [\M W_c]_{c=1}^{N_C}}$\; 
        \For{$k = 1$ \textbf{to} $K_{\mathrm{out}}$}
        \State $\V x_{k+1} = \mathrm{\textbf{FBS}}(\M L_k, \V x_k, k)$\;
        \State Compute $\M L_{k+1} = [\mathbf{diag} (\tilde{\M \Lambda}_c(\V x_{k+1})) \M W_{c}]_ {c=1}^{N_C}$\;
        \If{$ \norm{\V x_{k+1} - \V x_k}_2 < \epsilon_{\mathrm{out}} \norm{\V x_k}_2$}
        \State \textbf{break}
        \EndIf
        \EndFor
        \State \textbf{return} $\V x_{k+1}$ 
        \end{algorithmic}
        \end{algorithm}
    \fixme{Imposing Lipschitz constraints on the masks could potentially be helpful for proving the convergence of the fixed-point iterations.
    Note that, for our simple generator $\tilde{\boldsymbol{\Lambda}}$, we can efficiently enforce such constraints, as detailed in \cite{DGB2022}.
    For Theorem \ref{thm:convergnece} and \ref{thm:ExistFix}, we require the invertibility of the forward operator $\M H$ to define a single-valued update operator \smash{$T_{\tilde{\M \Lambda}, \M W, \V y}$}.
    For its set-valued generalization, which naturally arises if we drop the invertibility assumption, a stability analysis of the defining problem \eqref{eq:set_cvx_prblms_generic} was recently established in \cite{NeuAlt2024}.
    Note that in the single-valued case, such results are often key to establish the existence of fixed points.
    Independent of any theoretical considerations, we observed a converging behavior of both MMR and SAFI for the compressed-sensing MRI experiment in Section \ref{sec:NumericalResults}, where $\M H$ is not invertible. }
    
    \subsection{Parameterization of the Learnable Parameters}\label{sec:Parameterization}
    We now provide details of the parameterization for our two solution-adaptive regularizers.
    \fixme{The regularization strength $\lambda$ in \eqref{eq:inv_spc} and \eqref{eq:set_cvx_prblms_generic} is learnable for the corresponding reconstruction models.}
    For the \fixme{MMR} model \eqref{eq:inv_spc} from Section~\ref{sec:MajMin}, the remaining parameters are the linear operators $\{\M W$, $\M B\}$ and the concave potentials in $\M \Psi$.
    For the \fixme{SAFI} problem \eqref{eq:set_cvx_prblms_generic} from Section~\ref{sec:SolDrivAdapt}, the remaining parameters are the linear operators $\{\tilde{\M W}$, $\tilde{\M B}$, $\hat{\M B}\}$ and the activation functions $\{\boldsymbol \phi_1$, $\boldsymbol \phi_2$, $\boldsymbol \phi_3\}$.
    Taking a closer look at Algorithms~\ref{alg:FBS1}-\ref{alg:MajMin}, we observe that we actually only need access to $\M \Psi^\prime$ and not to $\M \Psi$ itself.
    Hence, we directly parameterize the derivatives $\M \Psi^\prime$ instead.
    
    \paragraph{Parameterization of Linear Operators}
    All linear operators are constructed with the \texttt{Conv2d} module from PyTorch.
    Here, we only detail the construction for the output dimension $N_C$.
    More specifically, we decompose each operator into $S$ stacked \texttt{Conv2d} modules; each with $N_C$ output channels, a kernel size $(k_s \times k_s)$, and a group size $G$.
    This was observed to be more effective than the direct use of a single \texttt{Conv2d} module with a larger kernel size \cite{GouNeuBoh2022,GouNeuUns2023}.
    Here, the group size $G$ controls the potential transfer of information across the different channels.
    In particular, if $G=1$, then each kernel of the $k$th layer, $k \in {2, \ldots, S}$, is convolved with all the ones of the $(k-1)$th layer.
    If $G = {N_C}$, then each kernel is only convolved with the one of its channel.

    \paragraph{Constrained Linear Operators}
    We impose constraints on some convolution kernels.
    All the kernels of $\M W$ and $\tilde{\M W}$ should have zero mean.
    To ensure this, let $\V w \in \R^{k_s^2}$ contain the vectorized elements of the respective kernel.
    Then, we can use the parameterization $\V w \mapsto (\V w -  (\V 1^{\top} \V w) / k_s^2)$, and optimize over unconstrained variables.
    For $\M B$, we impose that the kernel elements are positive and sum to one.
    Let $\V b \in \R^{k_s^2}$ be the vectorized kernel elements.
    Here, the implementation of the constraint is nonnegative with the parameterization $\V b \mapsto (\abs{\V b} (\V 1^{\top} \abs{\V b}))$.
    Note that $\abs{\cdot}$ is applied element-wise to $\V b$.

    \paragraph{Learnable Activation Functions}
        For the $\{\boldsymbol \phi_1$, $\boldsymbol \phi_2$, $\boldsymbol \phi_3\}$ in $\tilde{\M \Lambda}$, we rely on the learnable linear-spline framework introduced in \cite{bohra2021learning}.
        More precisely, we use a uniform grid centered at $0$ with stepsize $\Delta$ and $2M + 1$ points, $M \in \N$, and the B-spline of degree one defined as
        \begin{equation}
        \beta^{1}(x) = \begin{cases}1-|x|, & x \in[-1,1] \\
        0, & \text {otherwise.}\end{cases}
        \end{equation}
        Then, we parameterize each $\phi_{p,c}$ based on the vector $\mathbf{d}_{p,c} \in \mathbb{R}^{2M+1}$ of function values at the grid points as
        \begin{equation}\label{eq:SplinePara}
        \phi_{p,c}(x)= \begin{cases}d_{p,c,1}+\frac{d_{p,c,2}-d_{p,c,1}}{\Delta}\left(x+M \Delta \right), & x \in\left(-\infty, -M \Delta \right) \\
        \sum_{k=-M}^{M} d_{p,c,k + M + 1} \beta^1(x/\Delta-k), & x \in\left[-M \Delta, M \Delta\right] \\
        d_{p,c,2M + 1}+\frac{d_{p,c,2M + 1} - d_{p,c,2M}}{\Delta}\left(x-M \Delta\right), & x \in\left(M \Delta, \infty\right).\end{cases}
        \end{equation}
        In particular, $\phi_{p,c}$ is nonlinear on $[-M \Delta, M \Delta]$ and extrapolated linearly outside of this interval.
        
    \paragraph{Concave Potentials}
        For the \fixme{MMR} model, we parameterize the $\psi_c^\prime$, $c=1,\ldots,N_C$, as 
        \begin{equation}
            \psi_c^{\prime}(x) = \mathrm{clip}_{[0, 1]}\big(\sigma_c(r_c x)\big),
        \end{equation}
        where $\sigma_c \colon \R_{\geq 0} \to \R$ are learnable linear splines and $r_c \in \R_{>0}$ are learnable scaling constants that adapt the range.
        To parameterize $\{r_c\}_{c=1}^{N_C}$, we use the \texttt{nn.Parameter} module of \texttt{PyTorch}.
        To ensure their positivity, we use $\abs{r_c}$ instead of $r_c$ in the implementation. 
        As $\sigma_c$ is only defined on $\R_{\geq 0}$, we parameterize it with its $M+1$ values on the nonnegative part of the grid from \eqref{eq:SplinePara} denoted by $\V d_{c} \in \R^{M+1}$. 
        As $\psi_c$ must be concave, its derivative $\psi_c^{\prime}$ is constrained to be non-increasing on $\R$.
        This can be achieved by using a non-increasing $\sigma_c$ with $\sigma_c(0) = 1$.
        To enforce the condition $\sigma_c(0) = 1$, it suffices to fix $d_{c, 0} = 1$.
        Let $\M D \in \R^{M,M+1}$ be defined via $(\M D \V d_c)_{m-1} = (d_{c, m} - d_{c, m-1})$, $m =2, \ldots, M+1$.
        If all elements of $\M D \V d_c$ are non-positive, then $\sigma_c$ is non-increasing.
        To directly embed this constraint into the parameterization, we define 
        \begin{equation}
            \boldsymbol{P}_{\downarrow}(\V d_c) = \M S \textbf{clip}_{[-\infty, 0]} (\M D \V d_c) + \V 1_{M+1},
        \end{equation}
        where $\M S \in \R^{M+1,M}$ with $(\M S \V d_c)_m = \sum_{k=1}^{m-1} d_{c, k}$, $m = 1, \ldots, M+1$.
        By projecting the unconstrained coefficients $\V d_c \in \R^{M+1}$ to $\boldsymbol{P}_{\downarrow}(\V d_c)$, we ensure that the corresponding $\sigma_c$ is non-increasing. 
        With the proposed parameterization, the associated concave profiles $\psi_c \colon \R_{\geq 0} \to \R_{\geq 0}$ satisfy the following properties. 
        \begin{itemize}
            \item[i)] They are piecewise-quadratic, nonnegative, and increasing. 
            \item[ii)] We have that $0 \leq \psi^{\prime}_c(x)\leq 1$ for all $x \in \R_{\geq 0}$ and $\psi^{\prime}_c(0) = 1$. 
        \end{itemize}
    
        \section{Architecture and Training}\label{sec:Training}
        For our regularizers \eqref{eq:reg_rw} and \eqref{eq:RegConvMask}, we now describe the learning of the parameters detailed in Section~\ref{sec:Parameterization}.
        For both architectures and their respective reconstruction routines Algorithm~\ref{alg:MajMin} and~\ref{alg:FixIters}, we learn them by solving a denoising problem with additive white Gaussian noise of standard deviation $\sigma \in \{5/255, 15/255, 25/255\}$.
        Since the training procedure is exactly the same for both architectures, we restrict our discussion to \eqref{eq:reg_rw}.
        
        Let $\{\V x^m\}_{m=1}^M$ with $\V x^m \in \R^{40 \times 40}$ be a set of clean patches from the grayscale BSD500 dataset \cite{arbelaez_contour_2011}, and let $\{\V y^m\}_{m=1}^M=\{\V x^m + \V n_\sigma^m\}_{m=1}^M$ be some noisy versions, where $\V n_\sigma^m$ is a realization of the noise.
        For all experiments, we train with $M=238400$ patches.
        In the following, we collect all the learnable parameters of \eqref{eq:reg_rw} in the variable $\V \theta$.
        Given a noisy patch $\V y^m$, we obtain its denoised version $D^{n_1,n_2,n_3}_{\V \theta, \sigma}(\V y^m)$ by applying Algorithm~\ref{alg:MajMin}  with $K_{\mathrm{out}} = n_1$ steps.
        As discussed in Section~\ref{sec:MajMin}, fixing $K_{\mathrm{FBS}} = n_2 = 1$ for Algorithm~\ref{alg:FBS1} suffices to guarantee convergence in the denoising case.
        For calculating the involved $\prox_{\gamma \Vert \M  L \cdot \Vert_1}$ based on Algorithm~\ref{alg:2}, we use $K_{\mathrm{FBS}} = n_3$ steps.
        During the training phase, all the tolerances are set to $(-1)$ to ensure that the maximum number of steps is used.
        Now, we propose to learn the optimal parameters $\hat{\V\theta}$ in $D^{n_1,1,n_3}_{\V \theta, \sigma}$ based on the empirical risk
        \begin{equation}\label{eq:train_prblm}
            \hat{\V\theta} \in \argmin_{\V \theta} \sum_{n_1=4}^6 \sum_{n_3=10}^{12} \sum_{m=1}^{M} \bigl \Vert D^{n_1,1,n_3}_{\V \theta, \sigma}(\V  x^m + \V n_\sigma^m) - \V x^m \bigr\Vert_2^2.
        \end{equation}
        To solve \eqref{eq:train_prblm}, we use the ADAM optimizer \cite{KinJim2015} with a learning rate of $10^{-3}$ and a batch size of $128$ patches that are reconstructed for a single pair $(n_1,n_3)$.
        This pair is uniformly drawn at random from one of the possible values for each batch.
        As documented in \cite{AniAshKai2022}, using a random numbers of iterations has a regularizing effect when unrolling fixed-point iterations.
        In particular, this prevents the models from getting overfitted to a specific number of iterations.
        We perform $40$ training epochs and reduce the learning rate by a factor of $0.1$ at the $5$th and $10$th epoch.
        After each epoch, we evaluate the performance of the model on the Set12 validation data, and choose the output model as the one with the best performance.
        As a result, we obtain the regularization strength in \eqref{eq:inv_spc} as well as the linear layers and the potentials that appear in \eqref{eq:reg_rw}.
        \begin{remark}
            Instead of pursuing an unrolling approach for training, one can also aim to minimize \eqref{eq:train_prblm} for $n_1=n_3=\infty$ with implicit-differentiation techniques \cite{BKK2019,GilOngWil2021}.
            However, as already observed in \cite{GouNeuBoh2022}, it is usually unnecessary to fully compute the involved fixed points in $D_{\V \theta, \sigma}^{n_1,1,n_3}(\V y^m)$ to learn good parameters $\V \theta$ for the regularizer.
            Moreover, as we have two nested fixed-point problems, namely the problems \eqref{prb:gen_lasso} and \eqref{eq:prox_g}, this easily gets prohibitively expensive.
        \end{remark}
        
        \paragraph{Architecture and Initialization}
        \fixme{We use the default \texttt{nn.Conv2D} initialization for every linear layer, and initialize $\lambda$ as \num{1e-4}.
        Below, we discuss the remaining hyperparameters and initializations.}
        
        \textbf{\fixme{MMR} model}:
         For the operators $\{\M W_c\}_{c=1}^{N_C}$, we proceed as described in Section~\ref{sec:Parameterization} with $N_C = 64$, $S=2$, $k_s = 7$, and $G=1$.
         Further, we force the kernels to be zero-mean.
         For the modeling of $\{\M B_c\}_{c=1}^{N_C}$, we use a linear layer with $N_C = 64$, $S=2$, $k_s = 7$, and $G=64$.
         Here, we enforce that the kernels are positive and normalized. 
         For each concave potential $\psi_c$, we use $21$ gridpoints, which corresponds to $M=20$ and $\Delta=0.05 $.
         We initialize the expansion coefficients of the splines with zero, except $d_{c, 0} = 1$.
         Every $r_c$ is initially set to one. 
         
         \textbf{\fixme{SAFI} scheme}:
         To model the linear layers $\{\M W_{c}\}_{c=1}^{N_C}$, we choose $N_C = 64$, $S=2$, $k_s = 7$, and $G=1$.
         We enforce that the kernels are zero-mean.
         To model $\{\M W_{c, 1}\}_{c=1}^{N_C}$ , $\{\M W_{c, 2}\}_{c=1}^{N_C}$, and $\{\M W_{c, 3}\}_{c=1}^{N_C}$, we use linear layers with $N_C = 64$, $S=1$, $k_s = 7$ and $G=1$.
         Further, we use linear splines with no constraints to parameterize $\{\phi_{1, c}\}_{c=1}^{N_C}$, $\{\phi_{2, c}\}_{c=1}^{N_C}$, and $\{\phi_{3, c}\}_{c=1}^{N_C}$, as described in Section~\ref{sec:Parameterization}.
         For this, we use $21$ knots, which correspond to $M=10$ and $\Delta = 0.1$.
         We initialize all expansion coefficients of the splines with zero. 

        \paragraph{Fine Tuning} The interpretability of the learned denoiser $D^{n_1,1,n_3}_{\V \theta, \sigma}$ with the small $n_1$ and $n_2$ from the training stage is, however, limited.
        In particular, we only perform a partial minimization (unrolling) of \eqref{eq:inv_spc}.
        To remain within our theoretical setup, we need to iterate Algorithms~\ref{alg:FBS1}-\ref{alg:MajMin}  until convergence during the evaluation phase.
        Doing so without modifying the regularization strength $\lambda$ in \eqref{eq:inv_spc} has led to over-smoothing in our experiments.
        Moreover, the training of the model is for denoising only and not necessarily adapted to other inverse problems with $\M H \neq \textbf{Id}$.
        To deal with these issues, we propose to deploy \eqref{eq:reg_rw} with the previously learned parameters for \eqref{eq:inv_spc} and solely fine-tune $\lambda$ on a small set of task-specific validation data with a coarse-to-fine grid search, as described in \cite{GouNeuBoh2022}.
        Hence, we get two different denoisers for our numerical evaluation:
        first, the unrolled version $D^{n_1,1,n_3}_{\V \theta, \sigma}$, which is exactly what we have trained for, but which is not necessarily a fixed point of \eqref{eq:FixedPointIters};
        second, the \emph{exact} fixed point \smash{$D^{\infty,1,\infty}_{\V \theta, \sigma}$} with an adapted $\lambda$, which uses more iterations and for which our theoretical analysis holds.
        For the inverse problems, we use the parameters of the denoising models that are trained with $\sigma = 15 / 255$.
        Here, we only have the fixed-point-based reconstruction operator as we do not train for the task.

        \paragraph{Algorithm Hyperparameters for Evaluation} We aim to iterate Algorithms \ref{alg:FBS1}-\ref{alg:MajMin} until convergence, namely, up to machine precision.
        Still, we enforce an upper bound on the number of iterations for all algorithms.
        This ensures that we always remain within a reasonable computational budget.
        Independent of $\M H$, we set $K_{\mathrm{prox}} = 500$ and $K_{\mathrm{out}} = 10$. 
        For the denoising case with $\M H = \textbf{Id}$, we know that $K_{\mathrm{FBS}} = 1$ suffices for convergence.
        There, we use the iteration-dependent tolerance
        \begin{equation}
            f_{\epsilon,{\mathrm{prox}}}(k_{\mathrm{out}}, k_{\mathrm{FBS}}) =  \begin{cases}
                  \num{1e-3}(0.01)^{\frac{k_{\mathrm{out}}}{5}}, & k_{\mathrm{out}} \leq 5 \\
                  10^{-5}, & k_{\mathrm{out}} > 5.\\
            \end{cases}
        \end{equation}
        For $\M H \neq \textbf{Id}$, we set $K_{\mathrm{FBS}} = 1000$ and use the iteration-dependent tolerances
        \begin{equation}
             f_{\epsilon,\mathrm{FBS}} (k_{\mathrm{out}}) =  \begin{cases}
                  10^{-3} (0.01)^{\frac{k_{\mathrm{out}}}{5}}, & k_{\mathrm{out}} \leq 5 \\
                  10^{-5}, & k_{\mathrm{out}} > 5, \\
              \end{cases}
        \quad\text{and}\quad f_{\epsilon,{\mathrm{prox}}}(k_{\mathrm{out}}, k_{\mathrm{FBS}}) =  \begin{cases}
                  3\epsilon_{\mathrm{FBS}} (\frac{1}{9})^{\frac{k_{\mathrm{FBS}}}{50}}, & k_{\mathrm{FBS}} \leq 50 \\
                  \frac{\epsilon_{\mathrm{FBS}}}{3},  & k_{\mathrm{FBS }} > 50.\\
              \end{cases}
        \end{equation}
        To summarize, for efficiency, the inner subproblems are solved with lower precision early on, while the precision for the later stages is higher to ensure convergence.
        This is a common technique to accelerate majorization minimization models \cite{SunBabPal2017} and the FBS algorithm \cite{SchRouBac2011,VilSalBal2013}.

        \section{Numerical Results}\label{sec:NumericalResults}
        First, we present denoising results as this is our training problem.
        Then, we deploy the regularizers \eqref{eq:reg_rw} and \eqref{eq:RegConvMask}, which we learned for denoising, to a MRI problem without additional training.
        For this, we need to adapt the $\lambda$ in \eqref{eq:inv_spc} and \eqref{eq:set_cvx_prblms_generic} on some (small) validation set.
        With this task shift, we want to underline the universality of our approach. \fixme{The code for our experiments is available on GitHub\footnote{ \href{https://github.com/mehrsapo/MMR_SAFI}{https://github.com/mehrsapo/MMR\_SAFI}}. In this section, the images of each row in a figure are plotted with the same grayscale. }
        \subsection{Denoising}\label{sec:Denoising}
        \begin{figure}[hpt]
            \begin{center}
              \includegraphics[width=17cm,height=10cm,keepaspectratio]{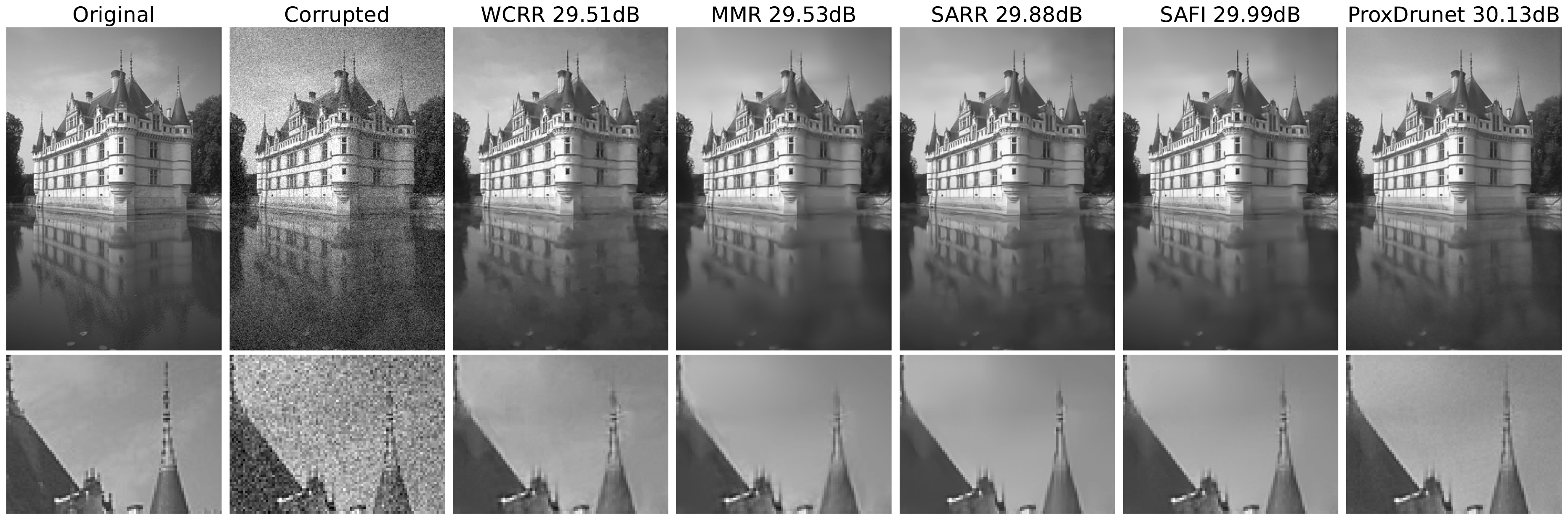}
              \caption{Denoising of the \emph{castle} image corrupted by additive white Gaussian noise with $\sigma=25/255$. }
              \label{fig:denoising_compare}
            \end{center}
            \vspace{.3cm}
            
            \begin{center}
              \includegraphics[width=17cm,height=10cm,keepaspectratio]{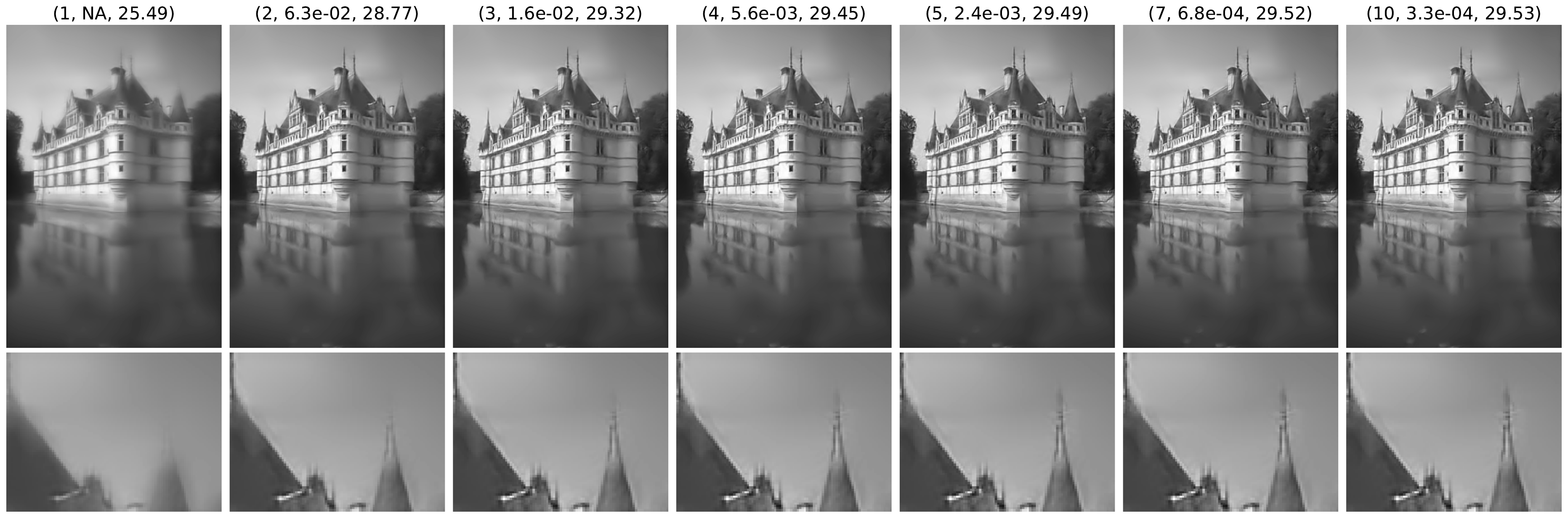}
              \caption{Solution path of the MMR method for denoising with $\sigma=25/255$.
              Each image ($k$, $e_k$, $\text{PSNR}_k$) represents $\V x_{k+1}$ at the $k$th step of Algorithm \ref{alg:MajMin}, with relative error $e_k = \frac{\norm{\V x_{k+1} - \V x_k}_2}{\norm{\V x_k}_2}$.}
              \label{fig:mmrr_denoising}
            \end{center}
            \vspace{.3cm}
            
            \begin{center}
             \includegraphics[width=17cm,height=10cm,keepaspectratio]{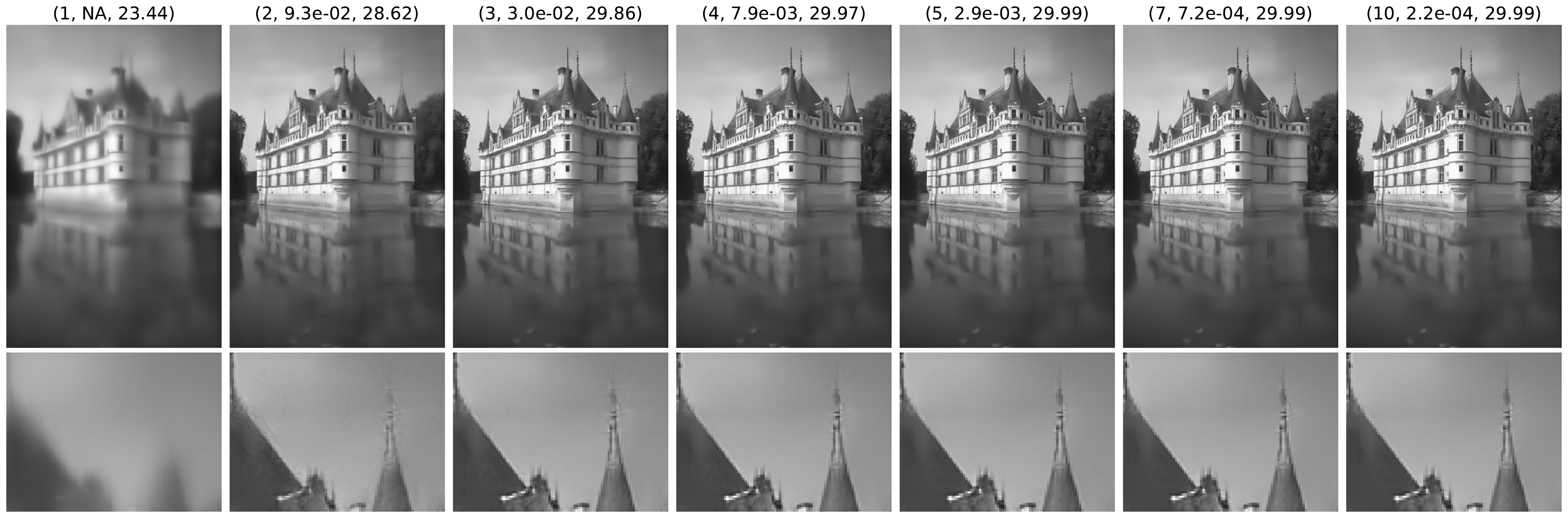}
             \caption{Solution path of the SAFI scheme for denoising with $\sigma=25/255$.
             Each image ($k$, $e_k$, $\text{PSNR}_k$) represents $\V x_{k+1}$ at the $k$th step of Algorithm \ref{alg:FixIters}, with relative error $e_k = \frac{\norm{\V x_{k+1} - \V x_k}_2}{\norm{\V x_k}_2}$.}
              \label{fig:safi_denoising}
            \end{center}
        \end{figure}
        Before investigating the qualitative behavior of the proposed regularizers \eqref{eq:reg_rw} and \eqref{eq:RegConvMask}, we first compare their quantitative performance with competing learned regularization methods.
        The achieved PSNR values on the BSD68 test set are given in Table~\ref{table:denoising_performance}. 
        \fixme{There, we compare our approach with BM3D, which is a popular baseline \cite{DabFoiKat2007}.
        We also compare with the WCRR model of \cite{GouNeuUns2023} and its spatially adaptive extension SARR \cite{NeuPouGou2023}, which both motivated our approach.
        Finally, we include Prox-DRUNet \cite{HurLec2022} as a regularizer with a deeper parameterization and some (loose) theoretical guarantees.}
        For the SAFI, we report results for both the training ($\text{SAFI}_5$) and the evaluation configuration (SAFI) with the $\lambda$ adaption.
        The performance difference between them is negligible.
        Hence, from now on, we solely use the evaluation configuration.
        Additionally, we provide a visual denoising comparison for the \emph{castle} image in Figure~\ref{fig:denoising_compare}.
        Here, we observe that our \fixme{SAFI scheme} recovers the tip of the spire, which is in general hard to achieve for $\sigma=25/255$.
        The spatial adaptivity helps to preserve sharp edges in the image.
        Still, all but the Prox-DRUNet method tend to slightly smooth the image.

        Regarding the qualitative behavior, we provide a solution path for MMR and SAFI in Figures~\ref{fig:mmrr_denoising} and \ref{fig:safi_denoising}, respectively.
        Somewhat surprisingly, the algorithm outputs a blurred reconstruction after the first step, in which all the noise is removed at the onset. 
        This initial reconstruction is then progressively sharpened throughout the remaining iterations.
        This behavior is particularly striking as SAFI still recovers the tip of the spire, see Figure~\ref{fig:denoising_compare}, which only reemerges in the later iterations.
        This is only possible since we update the mask iteratively  based on the previous reconstruction, which is not the case for the one step method SARR.
        As guaranteed by Theorem~\ref{thm:convergnece}, the residuals along the path for MMR in Figure~\ref{fig:mmrr_denoising} become small.
        The same is the case for SAFI where the iterates seem to converge to a fixed point, which necessarily exists due to Theorem~\ref{thm:ExistFix}.
        We observe the same converging behavior for all images in the BSD68 test set.
        Also, the visual behavior along the path is very similar in terms of an initial strong smoothing followed by a later recovery of sharp features. 

\begin{figure}[hpt]
            \begin{center}
             \includegraphics[width=17cm,height=10cm,keepaspectratio]{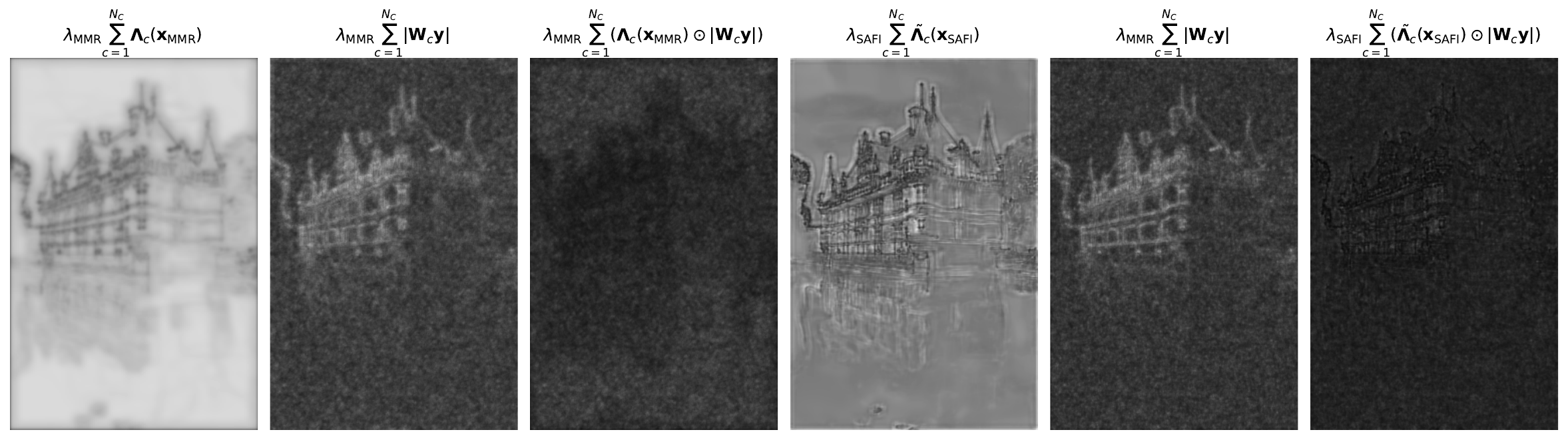}
             \caption{Masks and responses for the learned regularization architectures \eqref{eq:reg_rw} and \eqref{eq:RegConvMask}.
             Black corresponds to lower values and white to higher ones. \fixme{Note that $\{\M W_c\}_{c=1}^{N_C}$ is learned within the MMR and SAFI frameworks for the first and last three figures (from the left), respectively.}}
              \label{fig:masks}
            \end{center}
            \vspace{.3cm}

            \begin{center}
            \includegraphics[width=17cm,height=10cm,keepaspectratio]{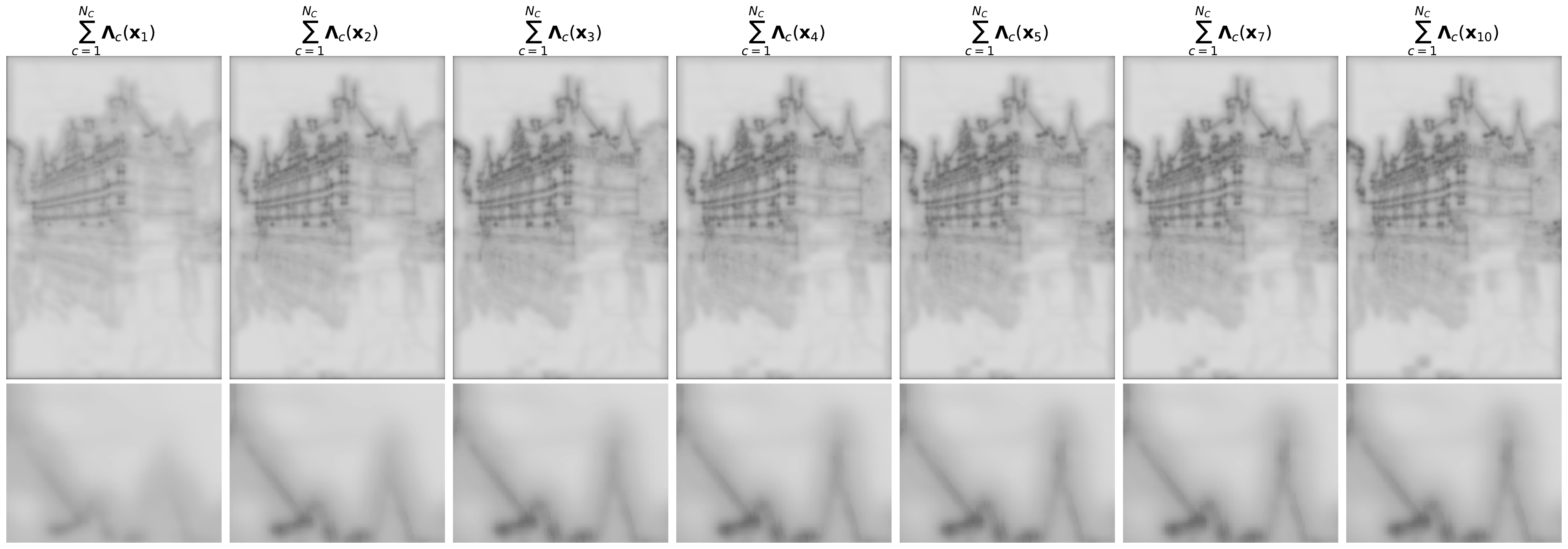}
            \vspace{.05cm}
            
            \includegraphics[width=17cm,height=10cm,keepaspectratio]{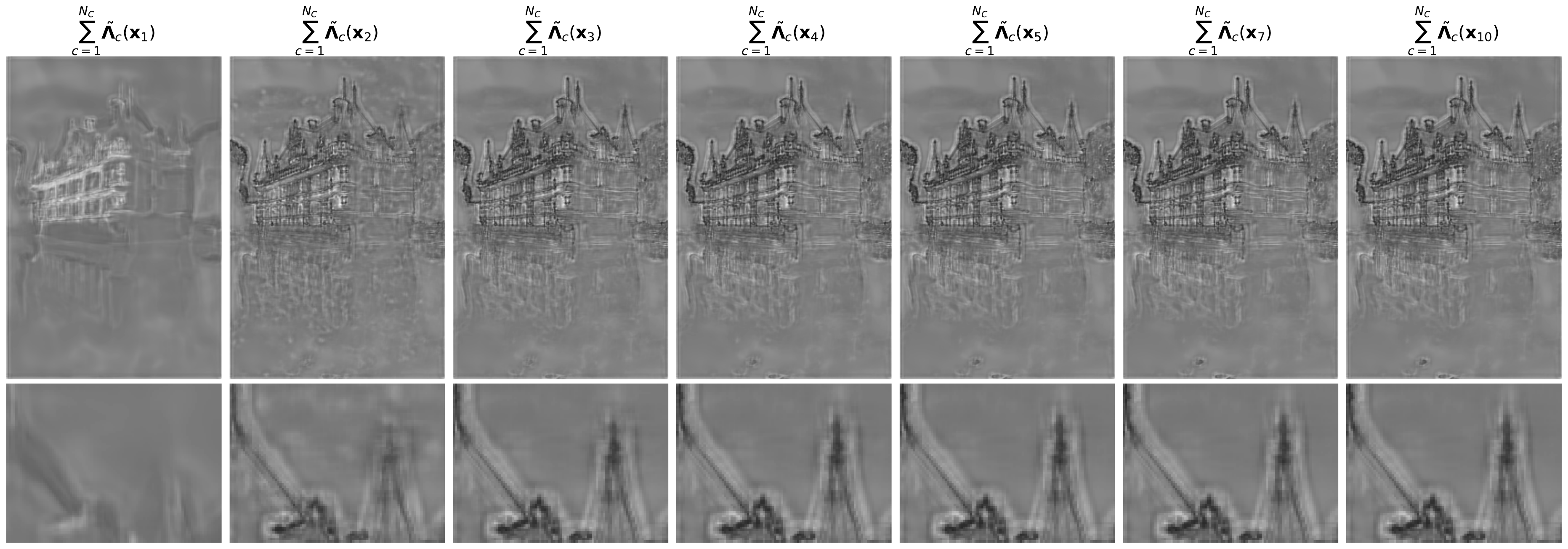}
            \caption{Evaluation of the masks for MMR (top) and SAFI (bottom).
            Both models become successively attentive to image structure.
            Still, the extracted structure in the MMR masks is far less pronounced.}
            \label{fig:safi_masks}
            \end{center}
        \end{figure}

        Now, we provide some intuition for the superiority of our approach over its nonadaptive counterpart.
        If $\norm{\M W \cdot}_1$ is a well-performing regularizer, the $\M W_c$ should not respond to the distinctive properties of an image.
        To investigate this for both MMR and SAFI, we display the response of the respective \smash{$\sum_{c=1}^{N_C} \abs{\M W_c \cdot}$} to the noisy image $\V y$ in Figure~\ref{fig:masks}.
        For both cases, the structure of the image is also triggered in addition to the noise.
        This leaves some room for improvements of the reconstruction results.
        In particular, we can dampen this undesirable response using the masks.
        Then, the effect of image structure on the regularization cost becomes less pronounced.
        In Figure \ref{fig:safi_masks}, we see how the masks become progressively more attentive to the image structure.
        Overall, the richer parameterization of the mask generator $\tilde {\boldsymbol \Lambda}$ for SAFI captures the image structure better.
        In particular, the masks for SAFI can still impose a high penalization in the vicinity of edges, whereas this is impossible for the masks from MMR.
        Overall, this results in a regularizer for which the image structures are less penalized.
        To conclude, the SAFI scheme leads to a better reconstruction performance than MMR model. 

        \begin{table}[p]
        \caption{Denoising performance (in terms of PSNR) on the BSD68 test set.
        \fixme{The average standard deviation of the PSNR for each image (based on 5 reconstructions) is similar for all settings and is roughly 0.02.}}
        \label{table:denoising_performance}
        \vspace{.1cm}
        \setlength\tabcolsep{6pt}
        \centering
        \begin{tabular}{lccccccc}
        \toprule
         Method & BM3D \cite{DabFoiKat2007} & WCRR \cite{GouNeuUns2023} & MMR & SARR \cite{NeuPouGou2023}  & SAFI & SAFI\textsubscript{5} & Prox-DRUNet \cite{HurLec2022}\\
        \midrule
        $\sigma=5/255$ & 37.54   & 37.6\fixme{5}  & 37.67 & 37.84   & 37.90 & \underline{37.91} & \textbf{37.97} \\
        $\sigma=15/255$ & 31.1\fixme{3}  & 31.2\fixme{0}  & 31.05 & 31.5\fixme{5}   & 31.5\fixme{6} & \underline{31.60}  & \textbf{31.70} \\
        $\sigma=25/255$ & 28.6\fixme{1}  & 28.6\fixme{8} & 28.\fixme{62}  & 29.07 & 29.0\fixme{5} & \underline{29.10}& \textbf{29.18}\\
        \bottomrule
        \end{tabular}
        \end{table}
        
        \subsection{Magnetic Resonance Imaging}\label{sec:MRI}
        \begin{table}[t]
        \centering
        \caption{PSNR (first columns) and SSIM (second columns) values for the MRI experiment.}%
        \label{table:reconstruction_performance_mri}
        \vspace{.1cm}
        \setlength\tabcolsep{2pt}
        \begin{tabular}{lcccccccc}
        \toprule
        & \multicolumn{4}{c}{4-fold single coil} & \multicolumn{4}{c}{8-fold multi-coil} \\
        & PD & PDFS & PD & PDFS & PD & PDFS & PD & PDFS\\
        \midrule
        Zero-fill ($\M H^{\top} \V y$) & 27.40 & 29.68 & 0.729 & 0.745 & 23.80 & 27.19 & 0.648 & 0.681  \\
        TV \cite{beck2009fast} & 32.44 & 32.67 & 0.833 & 0.781 & 32.77 & 33.38 & 0.850 & 0.824 \\
        CRR \cite{GouNeuBoh2022} & 33.99 & 33.75 & 0.880 & 0.831 & 34.29 & 34.50 & 0.881 & 0.852 \\
        WCRR \cite{GouNeuUns2023} & 35.78 & 34.63 & 0.899 & 0.838 & 35.57 & 35.16 & 0.894 & 0.856 \\
        SARR \cite{NeuPouGou2023} & \underline{36.25} & 34.77 & \underline{0.904} & 0.839 & \underline{35.98} & \underline{35.26} & \textbf{0.901} & \underline{0.858} \\
        Prox-DRUNet \cite{HurLec2022} & 36.20 & \textbf{35.05} & 0.901 & \textbf{0.847} & 35.78 & 35.12 & 0.894 & 0.857 \\
        MMR &  35.63 & 34.49 & 0.896 & 0.833 &  35.33&  34.97& 0.891 & 0.849 \\
        SAFI  & \textbf{36.43} & \underline{34.92}  & \textbf{0.908} &  \underline{0.844} & \textbf{36.06}  & \textbf{35.36} & \textbf{0.901} & \textbf{0.860} \\
        \bottomrule
        \end{tabular}
        \end{table}

        \begin{figure}[t]
            \begin{center}
\includegraphics[width=17cm,height=10cm,keepaspectratio]{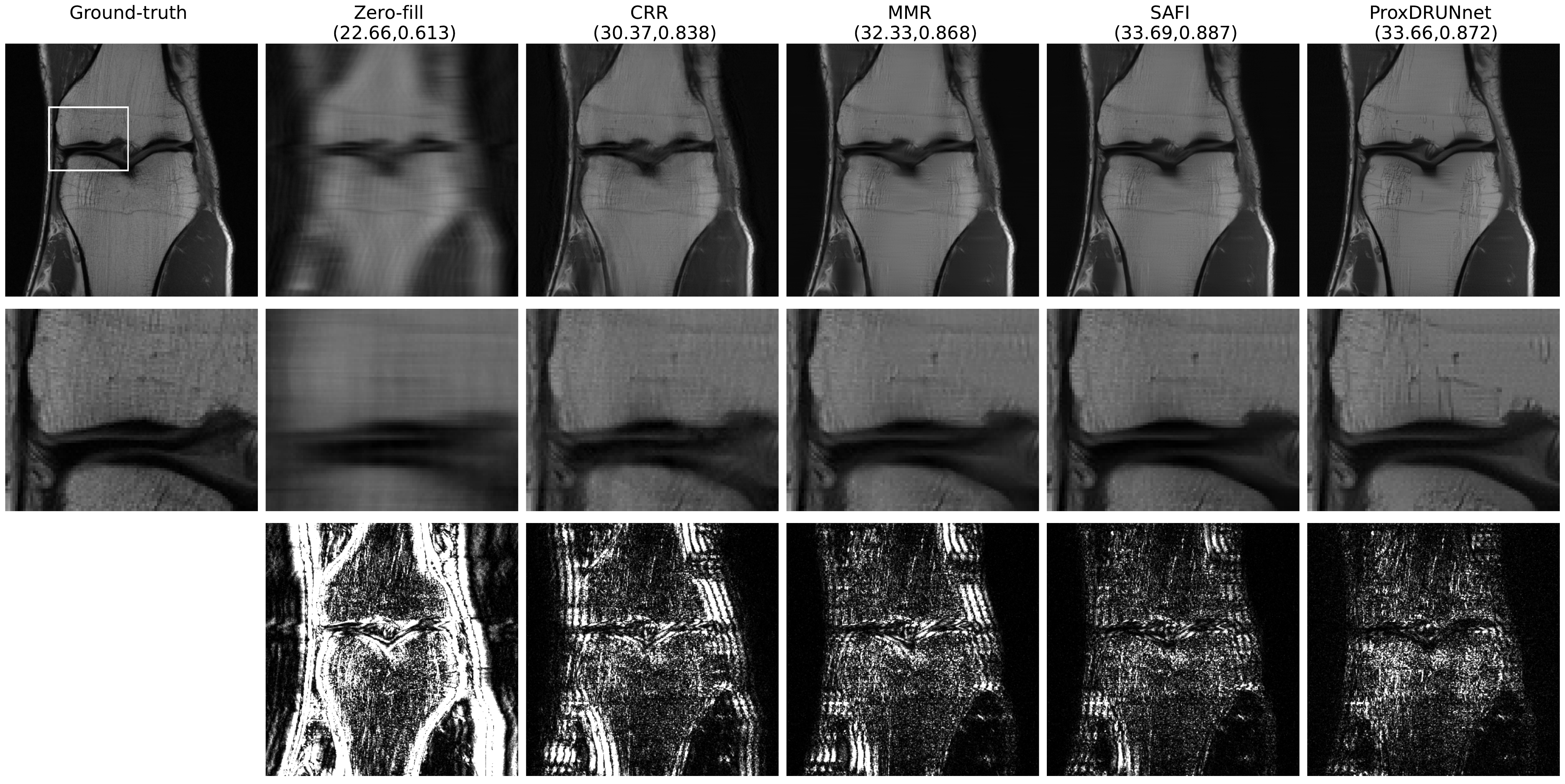}
              \caption{Reconstructions for multi-coil MRI (PD). The reported metric is (PSNR, SSIM). The second row contains the zoomed-in insets.
              The last row shows the squared value of the residuals\fixme{, which are cutoff at 0.003}.}
              \label{fig:mri}
            \end{center}
            \end{figure}

        \begin{figure}[t]
            \begin{center}
\includegraphics[width=17cm,height=10cm,keepaspectratio]{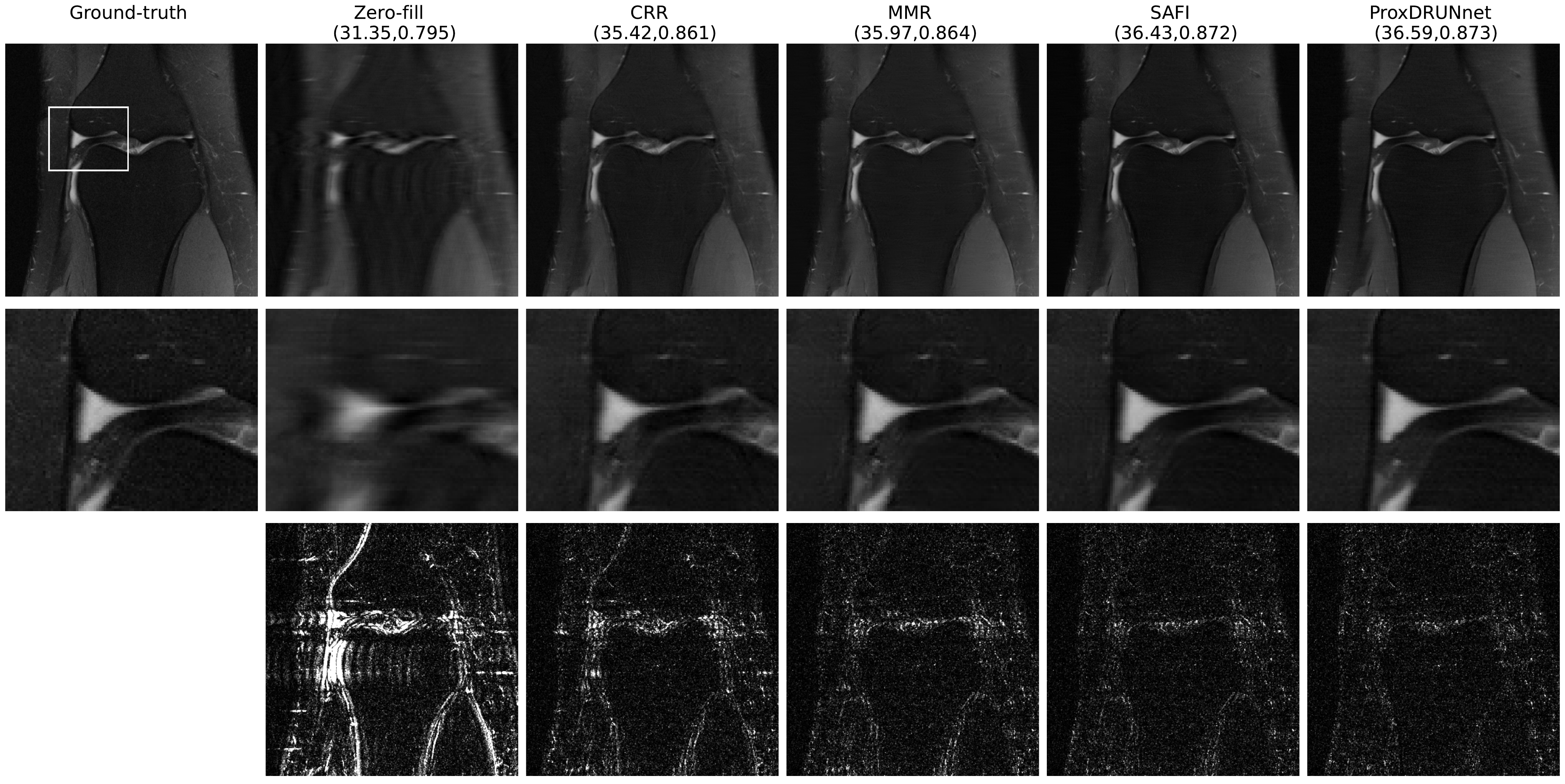}
              \caption{Reconstructions for single-coil MRI (PDFS). The reported metric is (PSNR, SSIM). The second row contains the zoomed-in insets.
              The last row shows the squared value of the residuals\fixme{, which are cutoff at 0.003}.}
              \label{fig:mri_pdfs}
            \end{center}
            \end{figure}
    
        Now, we deploy the proposed regularizers \eqref{eq:reg_rw} and \eqref{eq:RegConvMask} to solve MRI-reconstruction problems.
        We use the single- and 15-coil MRI setups detailed in \cite{GouNeuBoh2022}.
        For each setup, the ground-truth images consist of proton-density-weighted knee images from the fastMRI dataset \cite{zbontar2018fastMRI}, both with fat suppression (PDFS) and without fat suppression (PD).
        In total, this leads to four evaluation tasks.
        For each task, we use a validation set of ten images to fine-tune the regularization strength $\lambda$ in \eqref{eq:inv_spc} and \eqref{eq:set_cvx_prblms_generic}, respectively.
        We then report the test performance of the calibrated models on the remaining fifty test images. 
        To generate the ground-truth image, we use the fully sampled k-space measurements. 
        For the single-coil setup, we generate the measurements through a direct masking of the Fourier measures.
        In the 15-coil setup, we subsample the Fourier transforms of the ground-truth images multiplied by the respective sensitivity maps.
        For this, we use the BART \cite{uecker2013software} implementation of the ESPIRiT algorithm \cite{Uecker2014-uv}.
        The subsampling rate of each setup is determined by the acceleration factor $M_{\text{acc}}$ with the number of columns kept in the k-space being proportional to $1/M_{\text{acc}}$. Our single-coil setup is $4$-fold ($M_{\text{acc}} = 4$) and our multi-coil setup is $8$-fold ($M_{\text{acc}} = 8$).
        The measurements are then corrupted with additive white Gaussian noise of standard deviation  $\sigma = \num{2e-3}$.
        In Table~\ref{table:reconstruction_performance_mri}, we provide both the PSNR and structural-similarity index measure (SSIM) values on centered $(320\times 320)$ patches. \fixme{Here, we compare against the popular TV regularization,  the CRR as a state-of-the-art convex regularizer, its weakly convex extension WCRR, and the Prox-DRUNet as a popular PnP approach.
        Note that all of these methods are \emph{universal} in the sense that they can be deployed without additional training.}
        The full implementation details for the CRR and WCRR can be found in the respective papers.
        For Prox-DRUNet, we deploy the DRS-PnP algorithm proposed in \cite{HurLec2022}, which was previously adapted to our experimental setups in \cite{NeuPouGou2023}. 

        As we observe in Table \ref{table:reconstruction_performance_mri}, the MMR model achieves a performance close to that of the weakly convex model introduced in \cite{GouNeuUns2023}.
        This underlines again the strong relationship between the two regularization architectures and the associated models.
        The proposed SAFI regularizer achieves the best performance in three out of the four tasks and is second-best in the other one.
        Overall, these results indicate that our regularizers \eqref{eq:reg_rw} and \eqref{eq:RegConvMask} generalize well to inverse problems with the model parameters that were obtained by training on a denoising task.
        If enough data and compute resources are available, task-specific fine-tuning (second training stage) of the model parameters using the actual data or the forward operator $\M H$ can help to further increase the performance.
        
        In Figure \ref{fig:mri}, we provide multi-coil MRI reconstructions for a PD-type image.
        There, we observe that MMR results in a reconstruction that is sharper than the one with CRR, while the SAFI scheme yields even better results.
        Most importantly, these improvements do not come at the price of artifacts in the reconstruction.
        In terms of quantitative metrics, the Prox-DRUNet solution is comparable to SAFI.
        However, as we observe in the insets, this solution represents poorly the original texture of the images.
        In particular, it allows for sharp transitions but smooths out the textured parts of the image in this example.
        In Figure \ref{fig:mri_pdfs}, we investigate the single-coil setup for a PDFS image.
        This is the only case where the Prox-DRUNet is best on average.
        Although the Prox-DRUNet solution achieves higher PSNR than the SAFI solution, it is hard to observe pronounced visual differences between them.

        \subsection{Algorithmic Aspects for MMR and SAFI}
        \paragraph{Initialization}
        \fixme{
        In principle, the proposed MMR and SAFI reconstructions depend on the initialization of the schemes.
        The initializations are required to compute the first masks $\boldsymbol \Lambda$ and $\tilde{\boldsymbol \Lambda}$ for MMR and SAFI, respectively.
        In Algorithms \ref{alg:MajMin} and \ref{alg:FixIters}, we initialize with the solution of the nonadaptive convex problems \eqref{eq:set_cvx_prblms2} and \eqref{eq:set_cvx_prblms_generic} with $\boldsymbol \Lambda_c = \boldsymbol 1_N$ and $\tilde{\boldsymbol \Lambda}_c = \boldsymbol 1_N$, respectively. These plain experiments are denoted by CVX.
        To evaluate the sensitivity to this choice, we compare it against two alternatives.
        First, we perturb the proposed initialization by additive white Gaussian noise with $\sigma = 15/255$. These noisy experiments are denoted by Perturbed CVX.
        As an even stronger deviation, we use a random initialization, where each entry is drawn from the standard normal distribution. These challenging experiments are denoted by Random.
        The PSNR values of the respective reconstructions for the MRI experiment from Figure \ref{fig:mri_pdfs} are given in Table \ref{table:Robustness}.
        Most importantly, we observe that all variants eventually lead to the same PSNR.
        This indicates that the fixed point does not depend on the initialization.
        Unsurprisingly, the convergence to this fixed point occurs faster with a better initialization.
        We also observe the same behavior for other images.
        Finally, as indicated in Algorithms \ref{alg:FBS1} and \ref{alg:2}, the involved convex subproblems are always initialized with the minimizer of the previous one to accelerate the convergence.}
        \paragraph{Computational Complexity}
        
        \fixme{
        For the discussed MRI setups, the iterative SAFI approach is on average five times slower than the Prox-DRUNet approach, which does not incorporate any refinement steps.
        Reconstruction methods with a similar regularization architecture that do not incorporate a mask refinement (such as WCRR) can be even 50 times faster than SAFI.
        Memory-wise, SAFI has almost 10 times fewer parameters than ProxDRUNet and about 100 times more than WCRR.
        Since our approach brings valuable insights regarding weakly convex and spatially adaptive regularization, future work should focus on the improvement of the computational effectiveness of the approach.
        Since the subproblems \eqref{eq:set_cvx_prblms2} and \eqref{eq:set_cvx_prblms_generic} are convex, we can choose from a rich pool of methods for this goal.
        Moreover, we can draw from the literature on accelerating MM iterations \cite{HunLan2004,SunBabPal2017}.}
        
        \begin{table}[t]
        \caption{Robustness study: PSNR value after each MMR/SAFI update for the multi-coil MRI (PD) reconstruction experiment in Figure \ref{fig:mri} depending on the initialization.}
        \label{table:Robustness}
        \vspace{.1cm}
        \setlength\tabcolsep{6pt}
        \centering
        \begin{tabular}{lcccccccc}
        \toprule
         Update & 0 & 1 & 2 & 3  & 4 & 5 & 6 & Final\\
        \midrule
        MMR, CVX  & 34.24  & 35.93 & 36.30  & 36.38 & 36.41 & 36.43 & 36.43 & 36.44 \\
        MMR, Perturbed CVX  & 24.16 & 35.10 & 36.17 & 36.35 & 36.41 & 36.42 & 36.43 & 36.43 \\
        MMR, Random & 0 & 34.17 & 36.08 & 36.33 & 36.40 & 36.42 & 36.43 & 36.43 \\
        \midrule
        SAFI, CVX & 33.70 & 35.55 & 35.84 & 35.91 & 35.94 & 35.95 & 35.96 &  35.97 \\
        SAFI, Perturbed CVX & 24.10 & 35.09 & 35.76 & 35.90 & 35.94 & 35.95 & 35.96 & 35.97 \\
        SAFI, Random & 0 & 21.66 & 34.70 & 35.59 & 35.82 & 35.89 & 35.93 & 35.96\\
        \bottomrule
        \end{tabular}
        \end{table}

        \section{Conclusion}\label{sec:Conclusions}
        \fixme{We have proposed to use an iterative majorization-minimization regularization (MMR) along with solution-adaptive-fixed-point iterations (SAFI) as new families of data-driven regularizers. They give rise to a sequence of convex reconstruction problems. 
        Numerically, the minimizers associated with this sequence converged to a fixed point in all of our experiments.
        Overall, this leads to a robust, universal, and interpretable regularization method for inverse problems.
        A benefit of our simple mask generator for SAFI is that it is well-suited to the enforcement of Lipschitz constraints, which are in turn important to obtain stability estimates.
        Such constraints might be the key to the proof of the convergence of the fixed point iterations.
        Finally, it could also be interesting to explore other architectural constraints for generating the masks to obtain theoretical guarantees.}

        \section*{Acknowledgments}
        The research leading to this publication was supported by the European Research Council (ERC) under European Union’s Horizon 2020 (H2020), Grant Agreement - Project No 101020573 FunLearn, by the Swiss National Science Foundation, Grant 200020\_219356/1, and the German Research Foundation (DFG) within the SPP2298 under the project number 543939932.
        The authors thank Alexis Goujon for fruitful discussions.
        None of the authors have competing interests to disclose.

	\bibliography{refs,references,references_cvx}
    
\end{document}